\documentclass[aps,prb,twocolumn,superscriptaddress,floatfix,longbibliography]{revtex4-2}

\usepackage{graphicx} % Required for inserting images
\usepackage{booktabs}
\usepackage{siunitx}
\usepackage{physics}  % For physics symbols
\usepackage{amsmath}  % For additional math support
\usepackage{array}    % For better control of column alignment
\usepackage{subfigure} % For subfigures
\usepackage[utf8]{inputenc} 
\usepackage{braket}
\usepackage{comment}
\usepackage{hyperref}
\hypersetup{
    colorlinks=true,        % Enables colored links
    linkcolor=blue,         % Color for internal links
    citecolor=blue,         % Color for citations
    urlcolor=blue           % Color for URLs
}

\begin{document}

\title{Excitations and dynamical structure factor of $J_1-J_2$ spin-\(3/2\) and spin-\(5/2\) Heisenberg spin chains}

\author{Aman Sharma}
\email{a.sharma@epfl.ch}
\affiliation{Institute of Physics, EPFL, Lausanne,  Switzerland}

\author{Mithilesh Nayak}
\affiliation{Department of Physics, University of Fribourg, 1700 Fribourg, Switzerland}
\affiliation{Department of Physics and Astronomy, The University of Tennessee, Knoxville, Tennessee 37996, USA}

\author{Natalia Chepiga}

\affiliation{Kavli Institute of Nanoscience, Delft University of Technology, Lorentzweg 1, 2628 CJ Delft, The Netherlands}

\author{Fr\'ed\'eric Mila}
\affiliation{Institute of Physics, EPFL, Lausanne,  Switzerland}

%\author{Frédéric Mila} % Added accents in the name
%\affiliation{Institute of Physics, EPFL, Lausanne, Switzerland}

\date{\today}

\begin{abstract}
We study the dynamical structure factor of the frustrated spin-\(3/2\) \(J_1\)-\(J_2\) Heisenberg chains, with particular focus on the partially dimerized  phase that emerges between two Kosterlitz-Thouless transitions. Using a valence bond solid ansatz corroborated by density matrix renormalization group simulations, we investigate the nature of magnon and spinon excitations through the single-mode approximation. We show that the magnon develops an incommensurate dispersion at \(J_2 \approx 0.32J_1\), while the spinons, viewed as domain walls between degenerate valence bond solid states, become incommensurate at \(J_2 \approx 0.4J_1\) beyond the Lifshitz point (\(J_2 \approx 0.388J_1\)). The dynamical structure factor exhibits rich spectral features shaped by the interplay between these excitations, with magnons appearing as resonances embedded in the spinon continuum. The spinon gap shows a nonmonotonic behavior, reaching a peak near the center of the partially dimerized phase and closing at the boundaries, suggesting the appearance of a floating phase as a result of the condensation of incommensurate spinons. %, suggestive of a central charge \(c=2\) in the floating phase.
Comparative analysis with the spin-\(5/2\) case confirms the universality of these phenomena across half-integer higher-spin systems. Our results provide detailed insight into how fractionalization and incommensurate condensation govern the spectral properties of frustrated spin chains, offering a unified picture across different spin magnitudes.
\end{abstract}

\maketitle

\section{Introduction}

Quantum spin chains serve as fundamental models for studying strongly correlated systems and have been instrumental in revealing novel phases of matter, quantum criticality, and exotic excitations~\cite{giamarchi2003quantum, affleck1989quantum}. Among the simplest and most thoroughly explored is the \(J_1\)-\(J_2\) Heisenberg chain with spin-\(1/2\), which has provided deep insight into the role of frustration in one-dimensional systems~\cite{haldane1983continuum, majumdar1969next}. The model is defined by the Hamiltonian:
\begin{equation}
H = J_1 \sum_{i=1}^{N} \mathbf{S}_i \cdot \mathbf{S}_{i+1} + J_2 \sum_{i=1}^{N} \mathbf{S}_i \cdot \mathbf{S}_{i+2},
\end{equation}
where \(J_1\) and \(J_2\) are antiferromagnetic interactions between nearest and next-nearest neighbors, respectively. This model exhibits a rich phase diagram due to the competition between nearest-neighbor (\(J_1\)) and next-nearest-neighbor (\(J_2\)) antiferromagnetic interactions. At small \(J_2\), the system remains gapless and critical, described by the SU(2)\(_1\) Wess-Zumino-Witten (WZW) conformal field theory~\cite{eggert1996numerical,lavarelo2014spinon,ferrari2018dynamical}. As \(J_2\) increases beyond $J_2 \approx 0.2411J_1$ \cite{majumdar1969next}, the system undergoes a Berezinskii-Kosterlitz-Thouless (BKT) transition into a gapped dimerized phase with spontaneously broken translation symmetry and short-range correlations \cite{white1996dimerization, okamoto1992fluid}. The excitations in this regime fractionalize into spinons, and the spectrum becomes incommensurate for large enough \(J_2\), signaling a Lifshitz transition\cite{muller1981quantum, bougourzi1998exact, lake2005quantum}.

The spin-1 \(J_1\)-\(J_2\) chain exhibits a qualitatively different behavior\cite{chepiga2016dimerization}. At \(J_2=0\), the system lies in the celebrated Haldane phase, a gapped phase with hidden topological order and exponentially decaying correlations~\cite{haldane1983nonlinear, affleck1987rigorous,fath1993solitonic}. As \(J_2\) increases, the system remains gapped, but at a critical value of \(J_2 \approx 0.76J_1\), it undergoes a first-order phase transition into a next-nearest-neighbor Haldane phase~\cite{kolezhuk1996first, pixley2014frustration}. Remarkably, at the transition point, the lowest excitations are spinons, domain walls that separate regions of Haldane and NNN Haldane order. These spinons are deconfined and exhibit a narrow but incommensurate dispersion. Away from the transition, the spinons become confined into magnon bound states, and the spectrum evolves into that of conventional triplet excitations~\cite{sharma2025bound,vanderstraeten2020spinon}.

Motivated by these foundational studies, recent attention has turned to higher-spin models\cite{chepiga2020floating,chepiga20222}, where quantum fluctuations are reduced but frustration remains potent. In this regard, the spin-\(3/2\) \(J_1\)-\(J_2\) Heisenberg chain presents a particularly intriguing case due to its intermediate position between half-integer and higher-spin behaviors~\cite{chepiga2020floating, rachel2009spin}. Quantum fluctuations are still significant, yet the Hilbert space is rich enough to support complex valence bond structures. The model has garnered interest due to the interplay of frustration and quantum fluctuations, which gives rise to a rich phase diagram characterized by critical, dimerized, and floating phases \cite{chepiga2020floating}. For small values of \(J_2\), the system remains in a critical gapless phase described by the WZW SU(2)\(_1\) universality class. However, as \(J_2\) increases, the system undergoes a BKT transition to a partially dimerized gapped phase, where the translation symmetry is spontaneously broken. Upon further increasing \(J_2\), another BKT transition closes the spectral gap, leading to a floating phase with incommensurate correlations \cite{chepiga2020floating,roth1998frustrated}. Finally, at sufficiently large \(J_2\), the system undergoes a transition into a fully dimerized phase, akin to what has been observed in higher-spin systems \cite{chepiga20222}.

While significant efforts have been devoted to understanding the static properties of these phases\cite{chepiga2020floating,roth1998frustrated}, the nature of excitations and their contributions to the dynamical structure factor (DSF) remain less explored. The DSF is a crucial quantity that encodes information about spin dynamics and can be directly probed in inelastic neutron scattering experiments. Notably, the transition into incommensurate behavior and the interplay between magnon and spinon excitations in the partially dimerized phase raise intriguing questions about the spectral evolution across different regimes. While quantum fluctuations are comparatively weaker in spin-\(5/2\) systems, the phase diagram remains qualitatively similar, exhibiting a partially dimerized phase and transitions into incommensurate regimes \cite{chepiga20222}.

In this work, we investigate the dynamical properties of the partially dimerized phase in the spin-\(3/2\) \(J_1\)-\(J_2\) chain. Employing a valence bond solid (VBS) ansatz~\cite{affleck1988valence, niggemann1997quantum, tu2008valence}, validated against density matrix renormalization group (DMRG) calculations~\cite{white2004real, white2008spectral}, we analyze the dispersion relations of both magnon and spinon excitations. Using the single-mode approximation (SMA)~\cite{feynman1954atomic, bijl1940lowest, arovas1988extended, auerbach1998interacting}, we demonstrate that the magnons develop an incommensurate dispersion at \(J_2 \approx 0.32 J_1\), whereas spinons become incommensurate at a larger critical value, \(J_2 \approx 0.4 J_1\). This is consistent with the Lifshitz transition at \(J_2 \approx 0.388 J_1\), as identified in prior studies \cite{roth1998frustrated}. Moreover, we show that the gapless phases surrounding the partially dimerized phase can both be interpreted as the condensation of spinons. Finally, we turn to the spin-\(5/2\) \(J_1\)-\(J_2\) Heisenberg chain to examine the robustness of these features in higher-spin settings. Our comparative analysis reveals a consistent pattern in the evolution of magnon and spinon excitations across spin values in half-integer spin chains.

%Our results provide a deeper understanding of frustrated quantum magnetism, shedding light on the emergence of incommensurability in the DSF and its implications for experimental observations. Across both spin-\(\frac{3}{2}\) and spin-\(\frac{5}{2}\) cases, the spectral function is dominated by spinon continua, mirroring the spin-\(\frac{1}{2}\) case, and the magnon mode does not split out of the two-spinon continuum at any point in the Brillouin zone. This behavior contrasts with the spin-\(\frac{1}{2}\) Majumdar-Ghosh point at \(J_2/J_1 = 0.5\), where the magnon briefly detaches near \(k = \pi/2\). Furthermore, the SMA-derived spinon dispersion undergoes a transition to incommensurability as \(J_2\) increases. Two gap-closing spinon modes emerge within the partially dimerized phase, signaling the onset of the floating phase and suggesting the central charge to be 2 in this gapless phase. 

The rest of the paper is structured as follows. In Sec.~\ref{methods}, we describe the numerical methods and the construction of spinon and magnon excitations using SMA. Sec.~\ref{results} presents the DSF results and their comparison with SMA predictions. In Sec.~\ref{obsv}, we analyze the evolution of excitation spectra and incommensurability. Sec.~\ref{concl} concludes with a summary and outlook.

\begin{comment}
    The paper is organized as follows. In Sec.~\ref{methods}, we describe the numerical techniques employed in our analysis, including the time-dependent density matrix renormalization group (t-DMRG) for computing the dynamical structure factor, and the construction of magnon and spinon excitations using the single-mode approximation (SMA) on valence bond solid (VBS) states. In Sec.~III, we present the numerical results for the dynamical structure factor of the spin-\(\frac{3}{2}\) and spin-\(\frac{5}{2}\) \(J_1\)-\(J_2\) chains, along with a comparison to magnon and spinon dispersions obtained from SMA. Sec.~IV provides a detailed analysis of the evolution of spectral features—such as the spinon gap, incommensurate wavevectors, and the interplay between magnons and spinons—as a function of \(J_2\), supported by phenomenological observations. Finally, in Sec.~V, we summarize our findings and outline future directions for extending this framework to other frustrated or topological quantum systems.
\end{comment}

\section{Methods}
\label{methods}

\subsection{Time-dependent Density Matrix Renormalization Group}

To calculate the DSF of spin-3/2 and spin-5/2 Heisenberg chains with competing $J_1$ and $J_2$ interactions, we used the time-dependent density matrix renormalization group (tDMRG) method (sometimes also called TEBD, see Ref.~\cite{schollwock2011density} for terminology)
~\cite{white2004real,white2008spectral, white1992density, white1993density, schollwock2005density, schollwock2011density}. The DSF at zero temperature is defined as:
\begin{eqnarray}
S^{\alpha\alpha}(k,\omega)=\int dt\, e^{-i\omega t}\sum_{r_i,r_j}e^{ik(r_i-r_j)}\bra{\psi_0}S^\alpha_{r_i}(t)S^\alpha_{r_j}\ket{\psi_0},\nonumber
\end{eqnarray}
where $\alpha=x,y,z$, and $\ket{\psi_0}$ denotes the ground state. Physically, this DSF represents the Fourier transform of time-dependent spin-spin correlations and can be experimentally measured via inelastic neutron scattering (INS). In this work, since we are interested in a Hamiltonian invariant by rotation in spin space, we only consider the longitudinal DSF component, $S^{zz}(k,\omega)$.

The calculation of the DSF involves two main computational steps: (i) accurate determination of the ground state wavefunction, and (ii) real-time evolution of a state obtained by applying a spin operator to the ground state.

\subsubsection{Ground State Calculation}

The ground states of the spin-3/2 and spin-5/2 Heisenberg chains are computed using the standard DMRG method, employing a matrix product state (MPS) ansatz \cite{white1992density, white1993density, schollwock2005density, schollwock2011density}. We considered finite-size chains with 150 sites, using a fixed bond dimension \(\chi = 200\)–250 in the MPS representation, for spin-3/2. For spin-\(5/2\), we used chains of 90 sites that were simulated with \(\chi = 150\). Iterative sweeping with the one-site update algorithm continued until the variance per site reached a threshold of $10^{-3}-10^{-6}$ for spin-3/2 or $10^{-3}-10^{-5}$ for spin-5/2, ensuring a moderate to good accuracy of the ground states obtained for all parameter values of interest. Although fixed bond dimension DMRG may, in principle, risk convergence to metastable or locally optimal solutions, we verified the reliability of our ground states by performing additional convergence tests. Specifically, at selected representative points, we repeated calculations with larger bond dimensions and confirmed that the corresponding DSFs remained qualitatively and quantitatively unchanged (see Appendix~\ref{convergence}).

Our implementation is capable of handling much larger bond dimensions than those used in this study, allowing for high accuracy. However, larger bond dimensions significantly slow down the real-time evolution in tDMRG simulations, thereby significantly increasing the computational cost. As a result, we chose to work with moderately sized systems. In our simulations, we fix the bond dimension \(\chi\) throughout the tDMRG evolution, which ensures a well-controlled computational cost. The cost per time step scales as \(\mathcal{O}(\chi^3 d^3)\), where \(d = 2S + 1\) is the local Hilbert space dimension. Although larger bond dimensions could improve accuracy, we verified in Appendix~\ref{convergence} that our chosen \(\chi\) values yield converged results, balancing computational efficiency and spectral fidelity.

\subsubsection{Real-time evolution using tDMRG}

Following the computation of the ground state, the time evolution of the state produced by applying a local spin operator at the center of the chain was carried out using the tDMRG approach \cite{white2004real,white2008spectral}. The evolution operator \(U(t) = e^{-i t H}\) was approximated by decomposing it into Trotter gates~\cite{vidal2004efficient, daley2004time, feiguin2005time}. For the spin-3/2 and spin-5/2 chains studied in this work, we consistently utilized Trotter gates with a time step of \(0.05/J_1\), thus balancing computational efficiency and numerical accuracy. Throughout the evolution, the bond dimension of the resulting MPS was controlled by truncation, with the maximum allowed bond dimension being the same as that of the ground state. 

The application of the local spin operator initially creates a local excitation in the center of the chain. In time, spin-spin correlations spread across the chain. To avoid boundary artifacts, we terminated the simulation once the correlations approached the chain edges. The maximum evolution time was chosen accordingly, with typical values around \(t_f \approx 50/J_1\) to \(20/J_1\) for spin-3/2 chains of length \(N=150\), and \(t_f \approx 20/J_1\) to \(12/J_1\) for spin-5/2 chains of length \(N=90\). The velocity of this correlation spread depends on the spin magnitude and is generally faster for higher spin values. As a result, in the spin-\(5/2\) case, correlations reach the chain boundaries more quickly than in the spin-\(3/2\) case. 

\subsection{Numerical computation and extraction of DSF}

The DSF is obtained by means of a double Fourier transform of the calculated time-dependent spin-spin correlation functions. Due to simulations being performed for finite time intervals, numerical artifacts such as the 'ringing' effects naturally occur in the Fourier transform. To mitigate these artifacts, the correlation functions were multiplied by a Gaussian filter \(e^{-t^2/(2\sigma^2)}\) prior to performing the Fourier transform in time. We optimized the trade-off between energy resolution and numerical artifacts by choosing the Gaussian variance \(\sigma\) as \(0.276 t_f\).

The dynamical structure factor obtained from the t-DMRG method has the following Lehmann representation:
\begin{eqnarray}
   S^{zz}(k,\omega)= \frac{2\pi}{N}\sum_{\alpha} |\bra{\psi_\alpha} S^z_{-k}\ket{\psi_0}|^2\delta\left(\omega - \omega_\alpha(k)\right),
\end{eqnarray}
where \(\omega_\alpha(k)\) represents excitation energies above the ground state. Because our ground state is in the total spin-zero sector, applying the \(S^z\) operator excites states in the spin-1 sector. Thus, the spectral gap measured in the DSF corresponds to the lowest spin-1 excitation energy.

The computational complexity grows substantially with increasing spin values due to larger local Hilbert spaces and increased entanglement entropy. Despite these challenges, our methodology remains robust, achieving convergence in the ground state calculations and accurate real-time evolution for the spin-3/2 and spin-5/2 chains.

%Overall, the combination of DMRG and tDMRG used here proves highly effective for exploring the excitation spectra of higher-spin frustrated quantum spin chains, capturing both the continuum of modes and the discrete modes in the spectral functions.

\subsection{Ground state ansatz}

\begin{figure}[ht]
    \centering
    \includegraphics[width=\columnwidth]{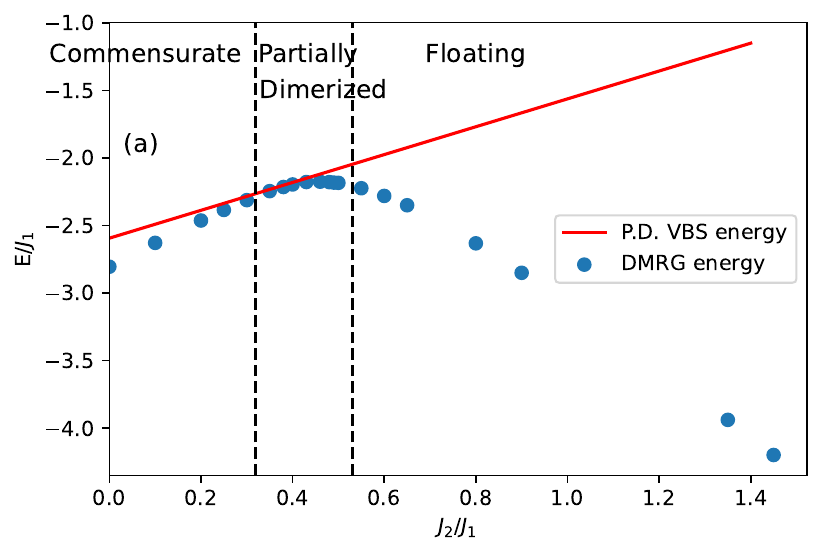}
    \includegraphics[width=\columnwidth]{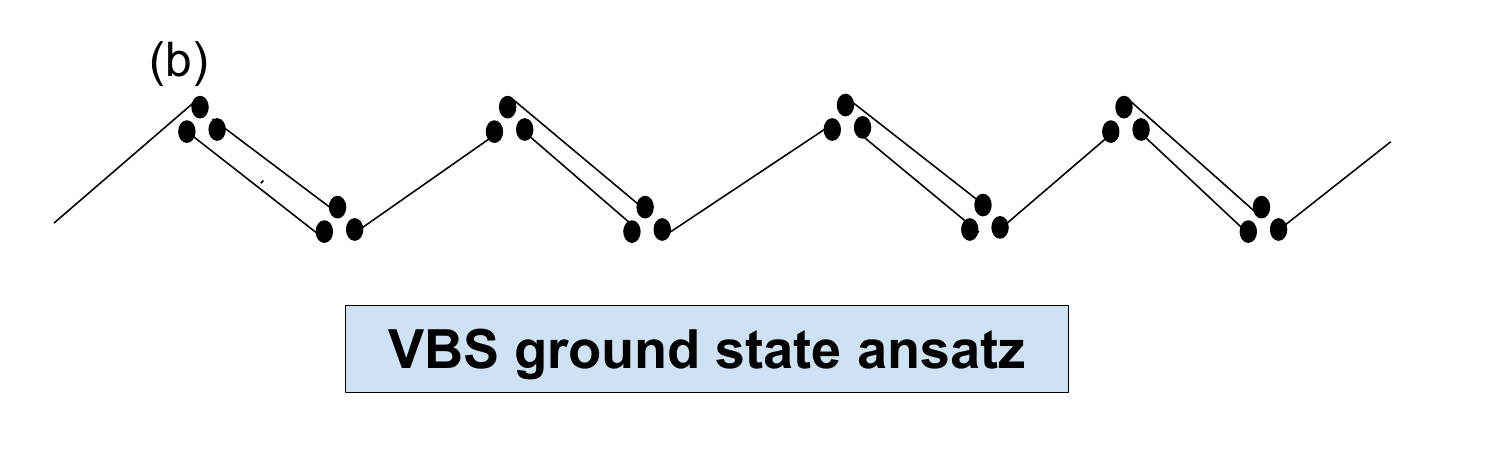}
    \includegraphics[width=\columnwidth]{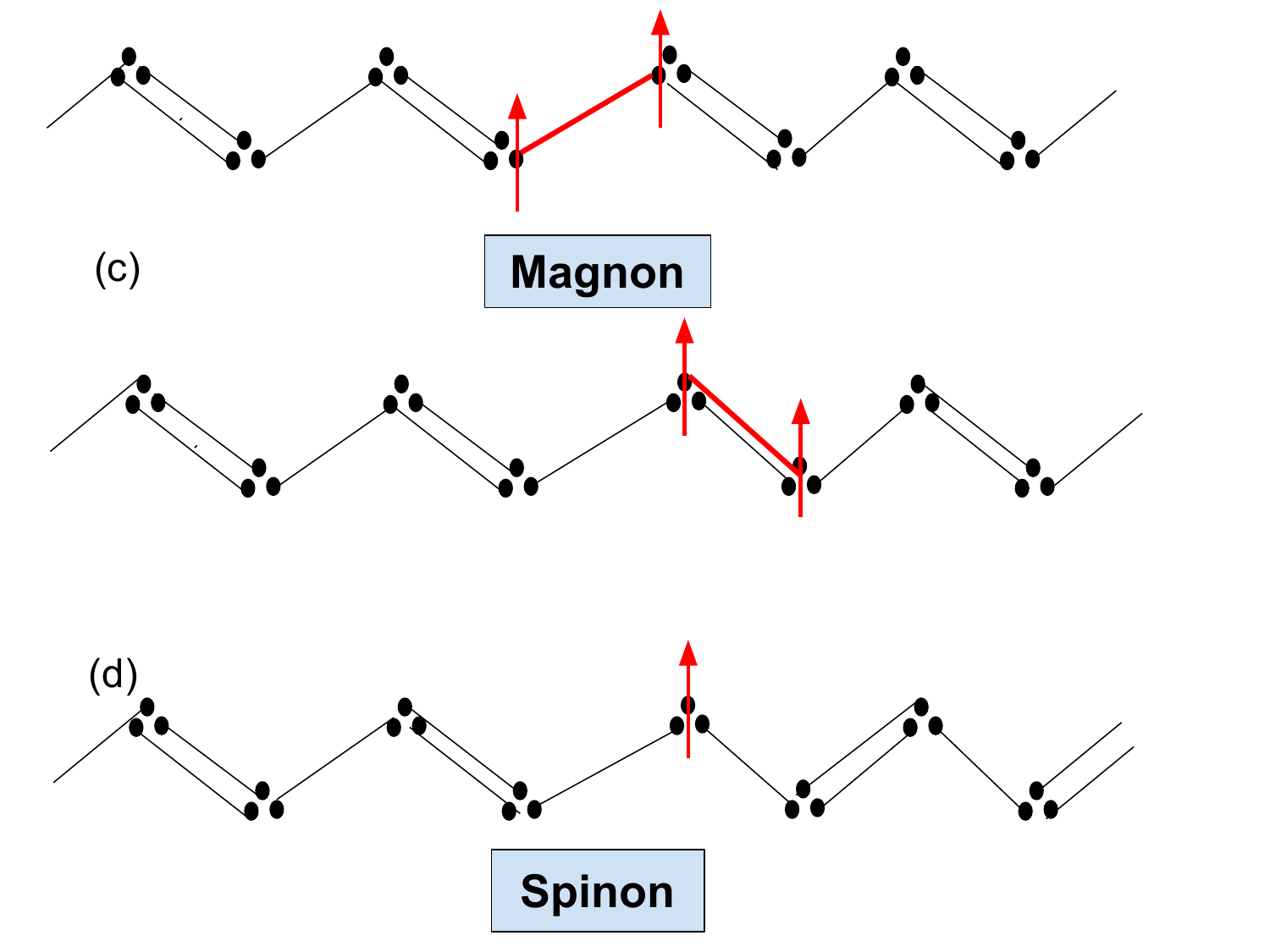}
    \caption{(a) Comparison of VBS and DMRG ground state energies (\(E/J_1\)), demonstrating the accuracy of the ansatz. (b) Illustration of the valence bond solid (VBS) ansatz for the spin-\(3/2\) \(J_1\)-\(J_2\) Heisenberg chain, showing one of the two degenerate ground states. $E$ denotes the ground state energy. (c) Construction of magnon excitations by promoting singlet bonds into triplet states on different bonds. (e) Spinon excitation as a domain wall between the two VBS states, which are degenerate in the thermodynamic limit. Spinon excitations are always produced in pairs. Only one of the two excitations has been depicted.}
    \label{fig:5}
\end{figure}

\begin{figure}[tb]
    \centering
    \includegraphics[width=\columnwidth]{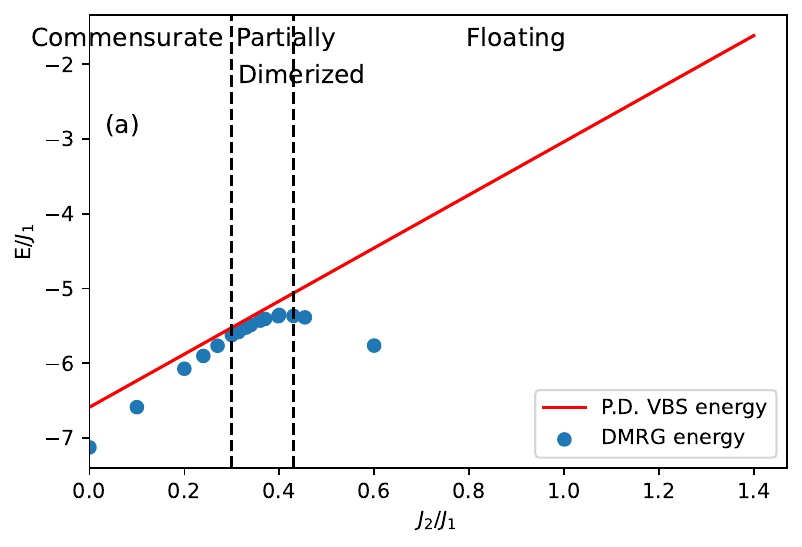}
    \includegraphics[width=\columnwidth]{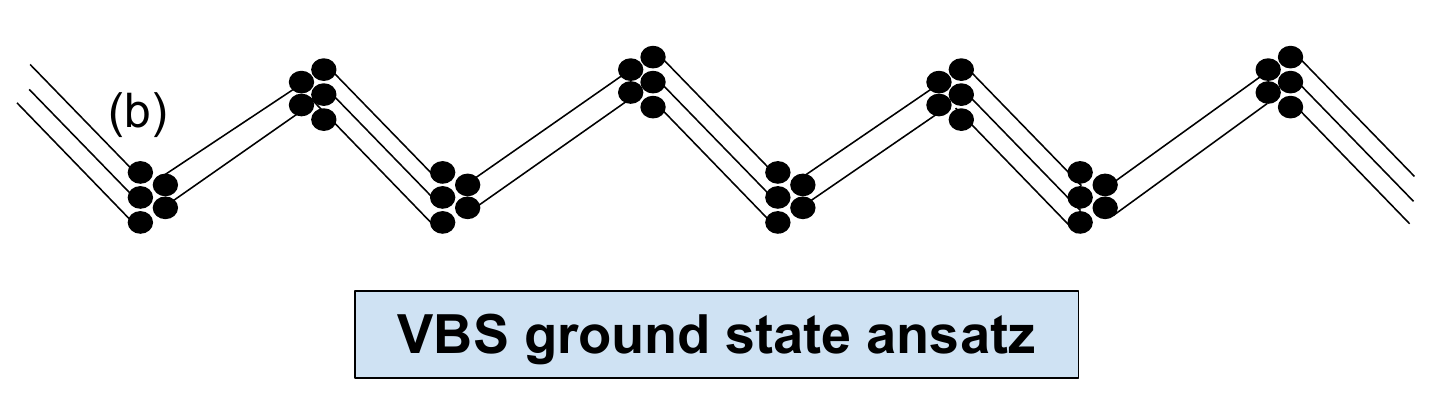}
    \includegraphics[width=\columnwidth]{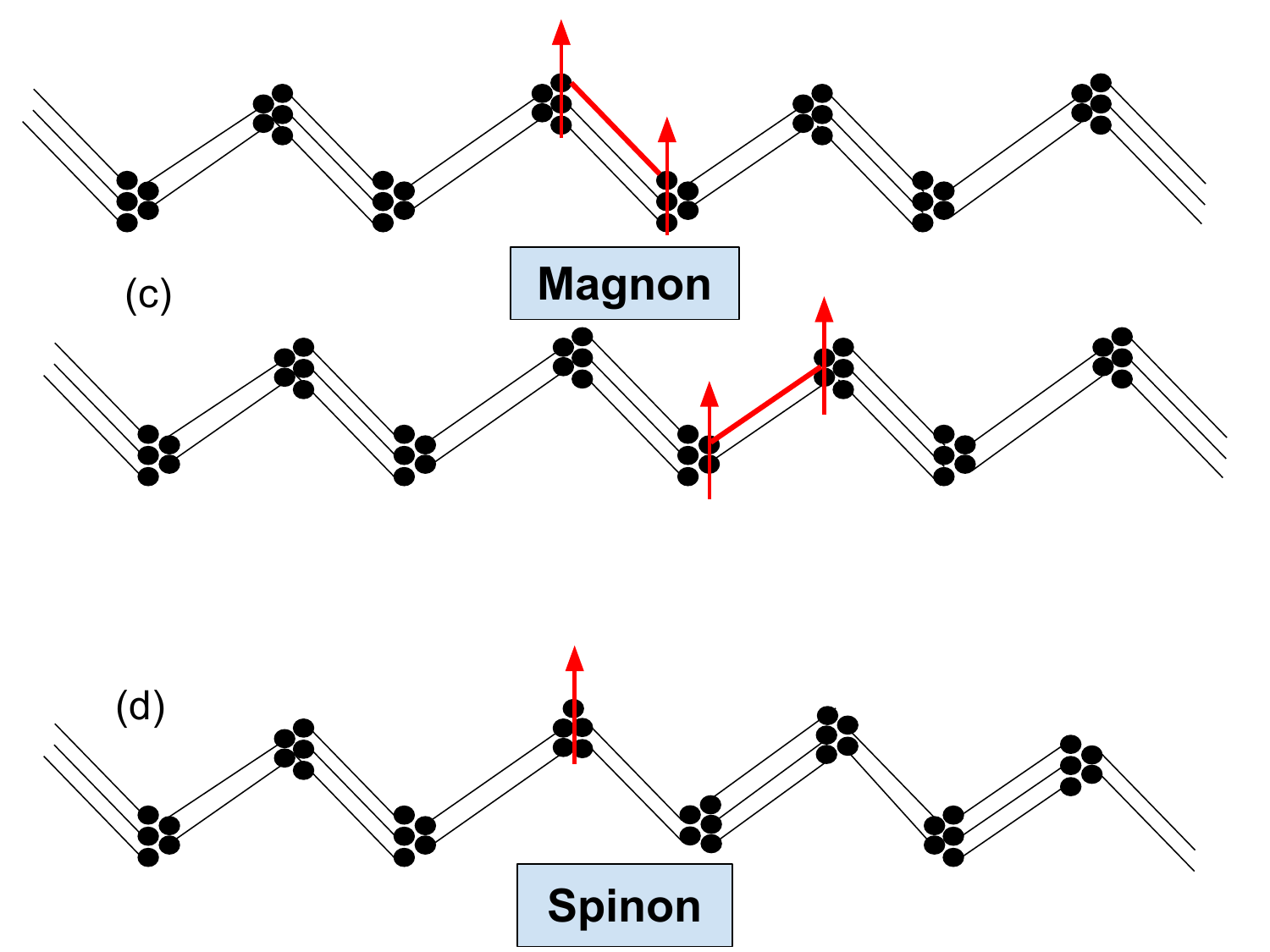}
    \caption{VBS ansatz and excitations for the spin-\(5/2\) \(J_1\)-\(J_2\) Heisenberg chain. Similar to Fig.~\ref{fig:5}, the VBS construction, magnon, and spinon states are shown, and the variational energies again match closely with DMRG calculations, validating the ansatz for higher-spin systems.}
     \label{fig:8}
\end{figure}

For the spin-\(3/2\) chain, our ground state ansatz for the partially dimerized phase in the valence bond picture is illustrated in Fig.~\ref{fig:5}b. Each site is modeled as comprising three symmetrically coupled spin-\(1/2\) particles. There exist two degenerate ground states, labeled GS1 and GS2. In GS1 (GS2), two singlet bonds form between neighboring sites on odd (even) bonds, and one singlet bond on even (odd) bonds.

Fig.~\ref{fig:5}a validates this ansatz by comparing the energy of the VBS state with the ground state energy of the \(J_1\)-\(J_2\) Heisenberg Hamiltonian obtained via DMRG simulations. The agreement between the two energies over the range \(J_2 / J_1 \in [0.3, 0.5]\) confirms the accuracy of the VBS ansatz in describing the ground state in this regime.

A similar picture holds for the spin-\(5/2\) case. The partially dimerized ground state ansatz now consists of alternating two and three singlet bonds on odd and even bonds, respectively, due to each site being composed of five symmetrically coupled spin-\(1/2\) particles. The degenerate counterpart features two and three singlet bonds on even and odd bonds, respectively (see Fig.~\ref{fig:8}b).

Fig.~\ref{fig:8}a confirms this ansatz by comparing the variational energy with the DMRG ground state results. The energies match closely in the vicinity of \(J_2 / J_1 \in [0.3, 0.4]\), again indicating the validity of the VBS construction in capturing the essential features of the partially dimerized phase in the spin-\(5/2\) chain. Numerical studies in Ref.~\cite{chepiga20222} indicate that a possible configuration like alternating one and four singlet bonds on even and odd bonds, respectively, is inconsistent with the local dimerization pattern observed in the true ground state of the spin-\(5/2\) \(J_1\)–\(J_2\) chain. This supports our chosen ansatz as the physically relevant one, while ruling out competing dimerization patterns.

\subsection{ Excitations with SMA}
\label{SMAsection}

Built on local variational excitations~\cite{girvin1986magneto, lake2013multispinon}, the SMA is a valuable theoretical approach widely used to calculate the dispersion relations of local excitations in quantum many-body systems~\cite{bijl1940lowest, feynman1954atomic, girvin1986magneto}. Originally developed in the context of quantum fluids, SMA was later successfully extended to quantum spin chains, most notably to describe magnon dispersion in VBS states such as the Affleck-Kennedy-Lieb-Tasaki (AKLT) model~\cite{arovas1988extended, auerbach1998interacting}. Given that the DMRG provides highly accurate ground states expressed naturally in MPS form, the SMA framework can be seamlessly generalized to MPS ground states to systematically compute the magnon and spinon dispersion relations.

For a local excitation with momentum \(k\) in a chain of length \(N\), the SMA state is defined as:
\begin{eqnarray}
\ket{k} = \frac{1}{\sqrt{N}}\sum_{j=1}^{N} e^{i k r_j}\,\Omega_j|\psi_0\rangle,
\label{Eq_sma_state}
\end{eqnarray}
where the operator \(\Omega_j\) creates a local perturbation (such as a bond flip from a singlet to a triplet) at bond (or site) \(j\) on the ground state \(\ket{\psi_0}\). The dispersion relation for this excitation, \(\omega(k)\), is obtained straightforwardly from:
\begin{eqnarray}
\omega (k)= \frac{\bra{k}H\ket{k}}{\braket{k|k}}-E_0,
\label{Eq_sma_defn}
\end{eqnarray}
where \(E_0\) is the energy of the ground state. Upon expanding the exponentials in Eq.~(\ref{Eq_sma_defn}) in terms of trigonometric functions, one obtains:
\begin{eqnarray}
\omega(k)=\frac{a_0+\sum_{n=1}^{N/2}a_n\cos(nk)}{1+\sum_{n=1}^{N/2}b_n\cos(nk)} - E_0,
\label{Eq_sma_defn_trig}
\end{eqnarray}
where \(a_n\) and \(b_n\) are parameters determined by matrix elements involving the Hamiltonian and overlaps between the states with the quasiparticle localized at different positions.

For the spin-\(3/2\) and spin-\(5/2\) \(J_1-J_2\) Heisenberg chains considered in this study, we employ SMA to systematically analyze two primary classes of excitations relevant to the interpretation of the DSF spectra.

\begin{enumerate}
\item {\it Magnon excitations:} In the partially dimerized phase, the ground state can be effectively represented as a VBS state consisting of one or two singlet bonds arranged periodically for the spin-\(3/2\) chain, and two or three singlet bonds for the spin-\(5/2\) chain. A local triplon excitation corresponds to promoting one of these singlet bonds to a triplet, as illustrated schematically in Fig.~\ref{fig:5}(c) for spin-\(3/2\) and in Fig.~\ref{fig:8}(c) for spin-\(5/2\). When such an excitation propagates coherently with a well-defined momentum \(k\), it gives rise to a magnon mode~\cite{auerbach1998interacting}. We construct explicit magnon states by introducing such local bond-flips into the partially dimerized VBS state. The resulting dispersion relations, computed via SMA, allow direct comparisons with numerically calculated DSF spectra, enabling the identification of distinct magnon features.

\item {\it Spinon (domain-wall) excitations:} Another fundamental class of excitations arises from domain walls separating two degenerate partially dimerized states, which are related by a one-site translation, as illustrated in Fig.~\ref{fig:5}(d) for spin-$3/2$ and Fig.~\ref{fig:8}(d) for spin-$5/2$. Unlike magnons, these domain wall excitations carry fractionalized spin quantum numbers (half-integers) and are therefore classified as spinons~\cite{lavarelo2014spinon}. Generating spinon states directly in an MPS framework involves non-local transformations, making analytical expressions difficult. Nevertheless, SMA allows for the numerical construction and systematic study of spinon dispersions by carefully preparing VBS states featuring domain walls. It is crucial to recognize that because of the dimerized nature of the state, spinons hop over every other lattice site; thus, care must be taken to properly account for this periodicity when performing the Fourier transform to momentum space. The analysis of spinon dispersions provides important insights into fractionalization phenomena near quantum phase transitions, which is particularly relevant for frustrated spin chains.
\end{enumerate}

Employing SMA to examine these two excitation types has allowed us to develop comprehensive interpretations of DSF features, deepening our understanding of the interplay between magnons and spinons in spin-\(3/2\) and spin-\(5/2\) Heisenberg chains.

Since the spinons are always produced in pairs and are free to move in the chain, there is a continuum in the spectral function. The continuum is characterized by the combined contributions of each spinon, and their total momentum and energy are given by:
\[
K = (k_1 + k_2)~\mathrm{mod}~2\pi, \quad E_K = E_{k_1} + E_{k_2},
\]
where \(k_1\) and \(k_2\) are the wavevectors of the individual spinons, and \(E_{k_1}\) and \(E_{k_2}\) are their respective energies.
	
In constructing the VBS ansatz for the excited states analyzed here, we adopted a matrix product state (MPS) based approach, as described in detail in Ref.~\cite{sharma2025bound}.

\section{Results}
\label{results}

\subsection{DSF of the spin-3/2 $J_1-J_2$ chains}
\label{sec:dsf_spin32_spin52}

\begin{figure*}[ht]
    \centering
    \includegraphics[width=0.9\textwidth]{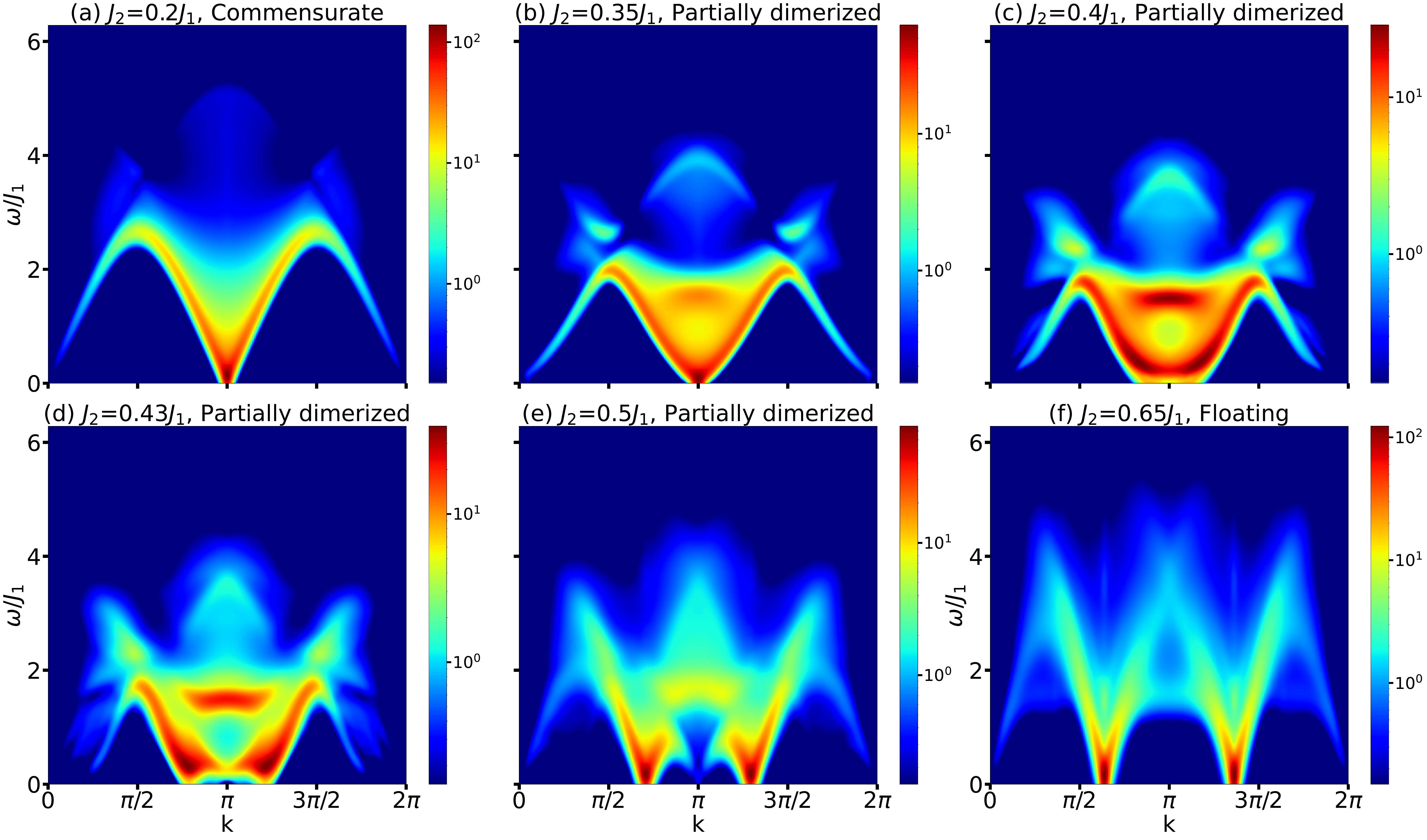}
    \caption{Dynamical structure factor \(S^{zz}(k,\omega)\) of the spin-\(3/2\) \(J_1\)-\(J_2\) chain at increasing values of \(J_2\). As \(J_2\) increases, the system transitions from a gapless critical phase (a) into a gapped partially dimerized phase (b-e), and reenters into a critical incommensurate phase (f). The emergence of incommensurate correlations is signaled by the splitting of low-energy spectral features.}
    \label{fig:dsf}
\end{figure*}

In this section, we report on the DSF of spin-$3/2$ $J_1$-$J_2$ chains using extensive numerical simulations via tDMRG. The DSFs for various values of the next-nearest neighbor coupling $J_2$ are presented in Fig.~\ref{fig:dsf}(a)-(f), providing a broad perspective on the evolution of the spectral functions across different phases.

At low values of $J_2$, specifically at $J_2=0.2J_1$, the spin-$3/2$ chain resides within a critical phase (C) characterized by gapless spin excitations. The DSF clearly exhibits a continuum with spectral boundaries consistent with the well-known de Cloiseaux-Pearson dispersion form\cite{des1962spin}, indicative of spinon-like fractional excitations. This continuum is typical for critical quantum spin chains and aligns with theoretical expectations of the spectra in a quantum critical phase.

Upon increasing $J_2$, the model transitions into a partially dimerized VBS phase, exhibiting a spectral gap. Within this phase, as $J_2$ further increases ($J_2=0.35J_1$ to $0.5J_1$), the DSF evolves into a richer structure with distinct features emerging at higher energies. Notably, within this intermediate parameter regime, the spin-spin correlation function becomes incommensurate, clearly reflected in the DSF as the low-energy excitation boundary develops two minima at incommensurate momenta. 

At still larger $J_2$ values (e.g., $J_2=0.65J_1$), the spin-$3/2$ system exits the dimerized phase and reenters another critical phase (floating phase), as evidenced by the return of a gapless spectrum reminiscent of the initial critical region. The DSF spectrum at this higher $J_2$ value looks like the superposition of two de Cloiseaux-Pearson continua\cite{des1962spin} touching zero energy at incommensurate wave-vectors. This is in line with the expectation that, for $J_1=0$, the DSF will be the superposition of two spin-3/2 DSF with a double unit cell, hence with a minimum at $\pi/2$.

%Our systematic study demonstrates how the DSF captures subtle yet critical changes in the underlying ground state and excitation spectrum.

%Additionally, the figure(Fig.~\ref{fig:6}(bottom-row)) compares the spectral weights with those of the two-magnon continuum, calculated using:
%\[
%K = k_{1} + k_{2}, \quad E_K = E_{k_1} + E_{k_2}.
%\]

%In Fig. \ref{fig:7}a, the two-spinon continuum is %compared with the spectral weights by superimposing %it on the DSFs. The spinon dispersion is displayed %below the DSF(Fig. \ref{fig:7}b).
\subsection{SMA analysis of spin-3/2 $J_1-J_2$ chains}
\label{sec:sma_analysis_spin32_52}

\begin{figure*}[ht]
    \centering
    % Subfigure 1
    \subfigure[]{
        \includegraphics[width=0.9\textwidth]{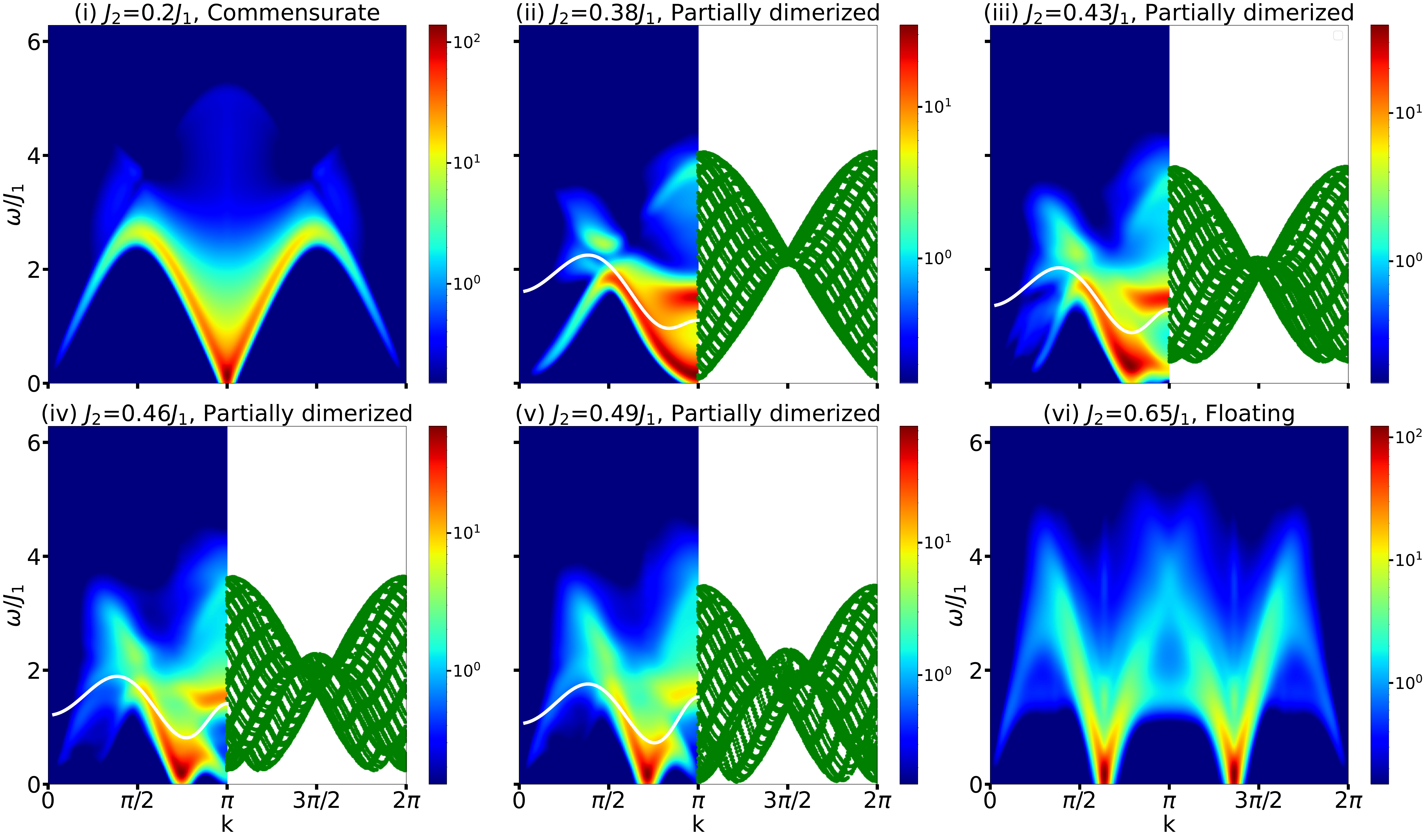}
        \label{}
    } \\
    % Subfigure 2
    \subfigure[]{
        \includegraphics[width=0.9\textwidth]{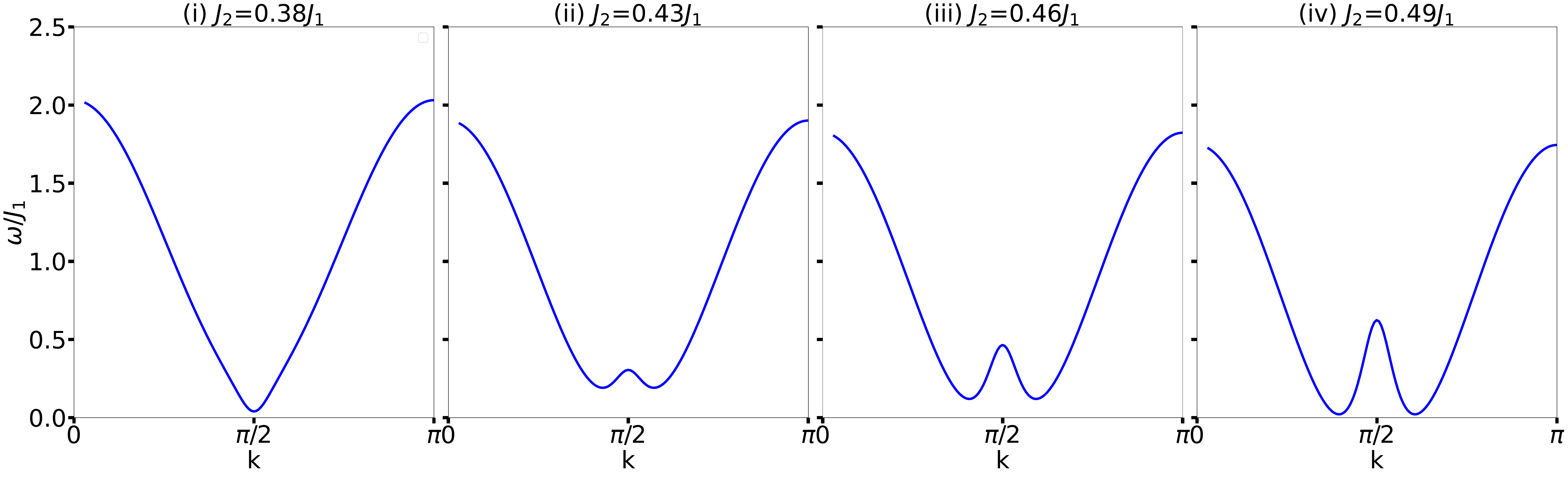}
        \label{}
    }
    \caption{(a) Dynamical structure factors for spin-$3/2$ chains compared with two-spinon continua (green shaded regions) calculated from the spinon dispersion of the domain-wall excitation between two degenerate partially dimerized states. The magnon dispersion (white solid curves), obtained via SMA on the partially dimerized state, is also shown. The SMA magnon dispersions pass through high spectral-intensity regions that are likely to be resonant magnons within the spinon continuum. (b) Corresponding spinon dispersions highlighting the splitting of dispersion minima as a clear sign of the onset of incommensurability, with the spinon gap showing non-monotonic behavior as $J_2$ is varied.}
     \label{fig:7}
\end{figure*}

To provide deeper insights into the spectral features observed in the DSFs of the spin-$3/2$ $J_1$-$J_2$ chain models, we conducted a comprehensive analysis using the SMA. Specifically, we focused on understanding the spectral contributions from spinon and magnon excitations within the partially dimerized phase.

We first examined the spinon excitations associated with domain walls separating two degenerate partially dimerized states, which differ by a translation of one lattice site. Using the SMA, we calculated the spinon dispersion and subsequently obtained the two-spinon continuum. In Fig.~\ref{fig:7}a, we compare the computed two-spinon continuum (shaded green region, right panel) to the corresponding DSF data (left panel). Remarkably, the spinon continuum accurately reproduces key features of the DSF spectra across a range of parameters, strongly suggesting that fractional spinon excitations play a dominant role in determining the spectral characteristics of the $J_1$-$J_2$ spin-$3/2$.

In parallel, we performed SMA calculations for magnon excitations on the partially dimerized ground state by explicitly breaking a singlet bond and converting it into a triplet bond, thus forming a local quasiparticle excitation. The resulting magnon dispersions (white solid curves) are also plotted on top of the DSF data in Fig.~\ref{fig:7}a. We observe that the magnon dispersion curves pass through several regions of high spectral intensity in the DSFs, indicating that magnon-like excitations are indeed present. However, given their embedding within the extensive spinon continuum, these magnons likely appear as resonant modes rather than stable, isolated quasiparticles, showing complex magnon-spinon interactions.

To further elucidate the behavior of spinon excitations, we present separately the calculated spinon dispersions in Fig.~\ref{fig:7}b. A notable feature of these dispersions is the progressive splitting of the dispersion minima as the system enters the incommensurate regime, clearly signaling the onset of incommensurate correlations. Additionally, the spinon gap initially increases, reaches a maximum, and subsequently decreases with increasing $J_2$, providing direct evidence of nontrivial spectral behavior near quantum phase transitions between the commensurate critical and incommensurate floating phase.

To gain a deeper theoretical understanding of the incommensurate features observed in the spinon dispersion, we reinterpret the SMA-derived dispersion in terms of an effective tight-binding model. In the SMA framework, the excitation energy is computed using Eq.\ref{Eq_sma_defn_trig}.
One can define \(\mathcal{O}_{ij} = \braket{\Omega_j | \Omega_i}\) as the overlap matrix, with \(|\Omega_j\rangle\) denoting a state with local spinon excitation at the site \(j\). Since these states are not orthogonal, the resulting subspace is nontrivial and must be handled carefully.

We now reinterpret the dispersion calculation as a variational problem within the subspace spanned by the non-orthogonal basis \(\{|\Omega_i\rangle\}\). The effective Hamiltonian \(\tilde{H}\), acting within this subspace, is defined as:
\begin{equation}
[\tilde{\mathcal{H}}]_{ij} = \sum_k \left[\mathcal{O}^{-1}\right]_{ik} \langle \Omega_k | \mathcal{H} | \Omega_j \rangle
\end{equation}
The generalized eigenvalue problem,
\begin{equation}
\sum_j \tilde{H}_{ij} \psi_j = \omega \sum_j \mathcal{O}_{ij} \psi_j,
\end{equation}
yields the spinon dispersion through the spectrum of \(\tilde{H}\). By performing a basis transformation using the inverse of the overlap matrix \(\mathcal{O}^{-1}\), this is equivalent to diagonalizing the projected Hamiltonian \(P H P\), where \(P\) projects onto the variational subspace.

In this representation, the energy dispersion can be written as:
\begin{align}
\omega(k) = \gamma_0 + \sum_{n=1}^{N/2-1} 2\gamma_n \cos(kn),
\label{eq:tight_binding_dispersion}
\end{align}
where the coefficients \(\gamma_n = \bra{\Omega_{N/2}} \tilde{H} \ket{\Omega_{N/2 + n}}\) can be interpreted as effective hopping amplitudes in a tight-binding model. Here, \(n\) denotes the distance between spinon positions. A more detailed exposition of this construction can be found in Refs.~\cite{lavarelo2014spinon, sharma2025bound}, where it was successfully applied to domain-wall excitations in spin-1/2 and spin-1 Heisenberg chains.

As noted previously in Section ~\ref{SMAsection}, spinon propagation occurs on every other site due to the underlying dimerized structure of the ground state. As a consequence, the effective hopping amplitudes $\gamma_n$ vanish for all odd $n$, and only even-$n$ contributions survive in Eq.~\ref{eq:tight_binding_dispersion}.

\begin{table}[ht]
\centering
\caption{Effective tight-binding hopping amplitudes $\gamma_n$ (in units of $10^{-2}$) extracted from SMA spinon dispersions for various values of $J_2/J_1$.}
\begin{tabular}{cccccc}
\toprule
$J_2/J_1$ & $\gamma_0$ & $\gamma_2$ & $\gamma_4$ & $\gamma_6$ & $\gamma_8$ \\
\midrule
% Example rows
0.38 & 111.875 & 47.0139 & -2.31481 & 1.54321 & -1.02881\\
0.43 & 97.8125 & 44.0625 & 3.47222 & -2.31481 &  1.54321\\
0.49 & 80.9375 & 40.5208 &  10.4167 & -6.94444 & 4.62962\\
% Add more rows as needed
\bottomrule
\end{tabular}
\label{tab:gammas}
\end{table}

We have computed the leading nonzero $\gamma_n$ values numerically for several values of $J_2/J_1$, as shown in Table~\ref{tab:gammas}. At $J_2/J_1 = 0.38$, the spinon dispersion exhibits a single minimum at $k = \pi/2$, indicating commensurate behavior. In this case, the second and fourth neighbor hoppings ($\gamma_2$, $\gamma_4$) exhibit alternating signs, with $\gamma_4$ being small and negative. This sign alternation ensures that the different cosine terms in Eq.~\ref{eq:tight_binding_dispersion} reinforce the same momentum minimum, stabilizing the commensurate dispersion.

However, as $J_2/J_1$ increases, $\gamma_4$ becomes positive while $\gamma_2$ remains positive, leading to constructive interference of multiple cosine terms in Eq.~\ref{eq:tight_binding_dispersion}. This competition between hopping amplitudes introduces a double-minimum structure in the spinon dispersion, manifesting as incommensurate modes. At $J_2/J_1 = 0.43$, a weak splitting appears, with $\gamma_4$ approximately one-tenth of $\gamma_2$. This splitting becomes more pronounced at $J_2/J_1 = 0.49$, where $\gamma_4$ grows to roughly one-fourth of $\gamma_2$, enhancing the competition and pushing the minima further away from $k = \pi/2$.

This tight-binding reinterpretation of the SMA dispersion makes the mechanism behind the incommensurability explicit: the interplay between second- and higher-neighbor hopping amplitudes destabilizes the commensurate minimum and leads to the emergence of incommensurate excitations.

\subsection{DSF and SMA analysis of the spin-5/2 $J_1-J_2$ Models}

% Second Figure (ii)
\begin{figure*}[ht]
    \centering
    \includegraphics[width=0.9\textwidth]{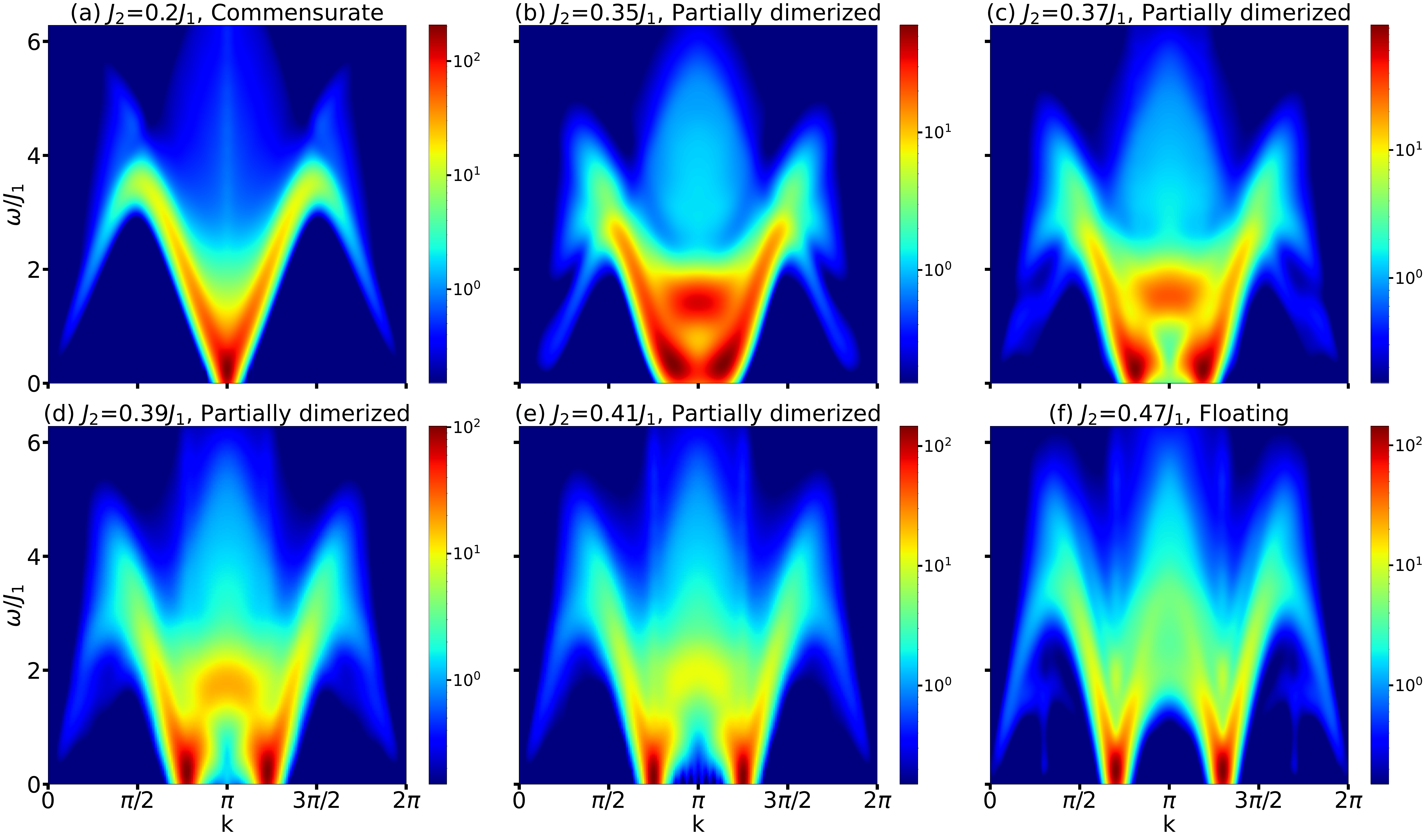}
    \caption{Dynamical structure factor \(S^{zz}(k,\omega)\) of the spin-\(5/2\) \(J_1\)-\(J_2\) Heisenberg chain at increasing values of \(J_2\). As \(J_2\) increases, the system transitions from a gapless commensurate phase (a) into a gapped partially dimerized regime (b–e), and then into a gapless incommensurate (floating) phase (f). The onset of incommensurate correlations is visible in the splitting and broadening of low-energy spectral features.}

    \label{fig:dsfb}
\end{figure*}

\begin{figure*}[ht]
    \centering
    % Subfigure 1
    \subfigure[]{
        \includegraphics[width=0.9\textwidth]{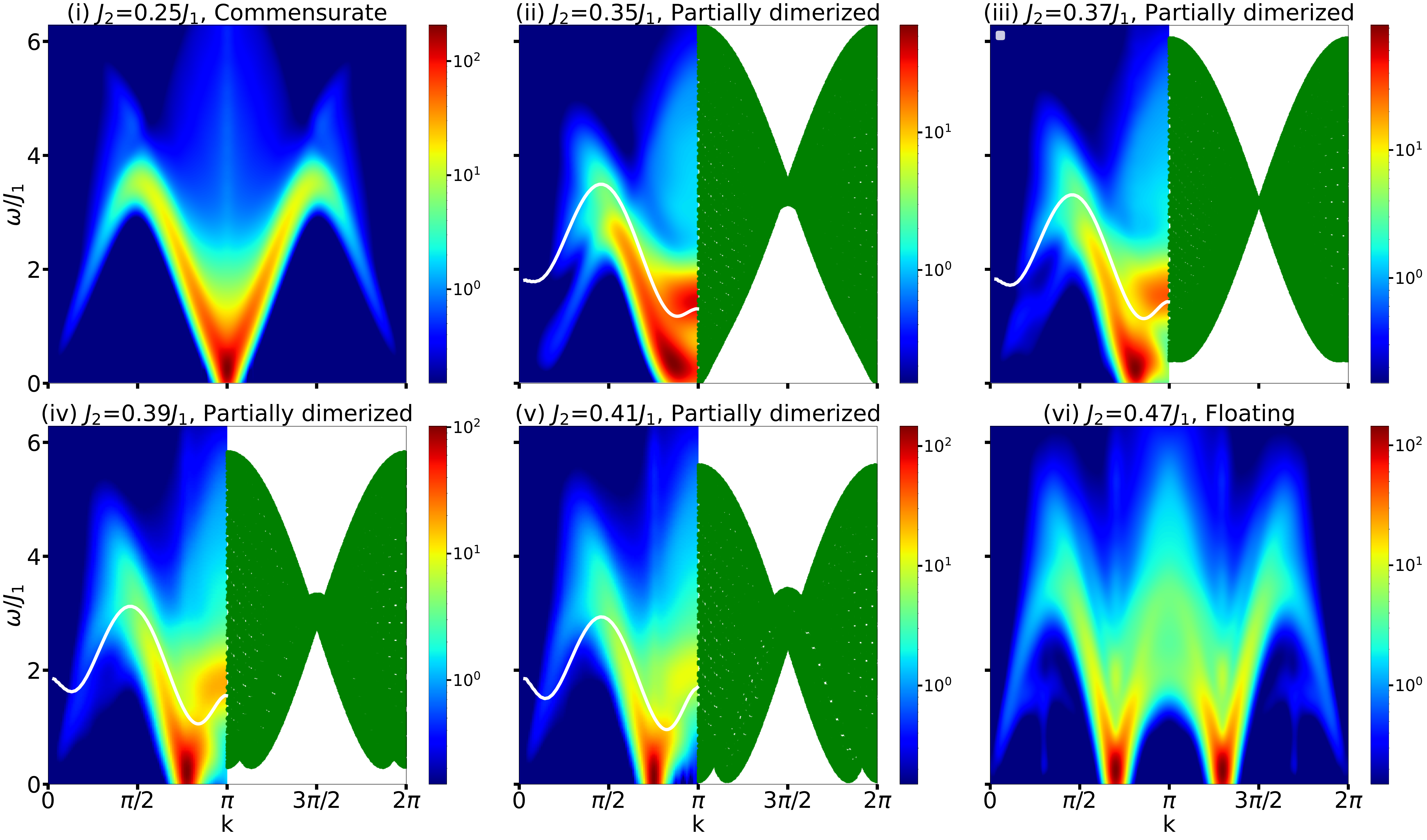}
        \label{}
    } \\
    % Subfigure 2
    \subfigure[]{
        \includegraphics[width=0.9\textwidth]{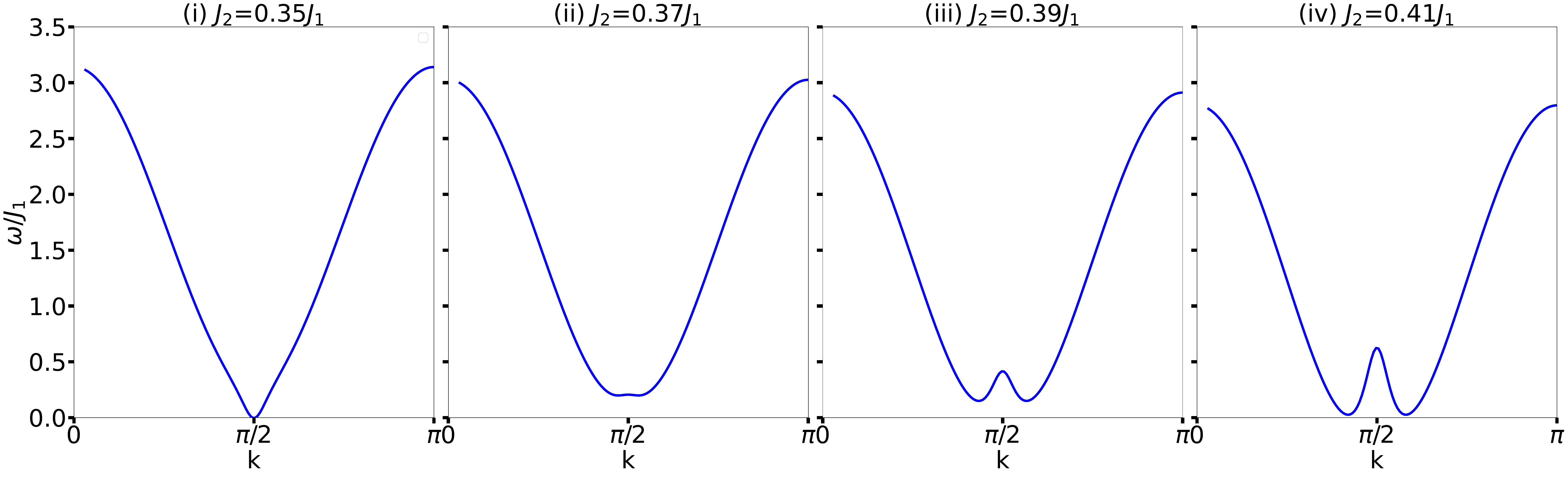}
        \label{}
    }
    \caption{(a) Dynamical structure factors of the spin-\(5/2\) \(J_1\)-\(J_2\) chain compared with two-spinon continua (green) and SMA-derived magnon dispersions (solid lines). Magnons trace regions of high spectral weight. (b) Spinon dispersions exhibit incommensurate minima and non-monotonic gap variation with \(J_2\).}
    \label{fig:10}
\end{figure*}

We now extend our study of the DSF to the spin-\(5/2\) \(J_1\)–\(J_2\) Heisenberg chain, using tDMRG simulations and SMA analysis. Figure~\ref{fig:dsfb} presents the DSFs for increasing values of \(J_2\). It captures the evolution of the spectral response as the system transitions through distinct quantum phases.

At small \(J_2\), the system exhibits a gapless commensurate critical phase and, as for spin-\(3/2\), the spectral functions are reminiscent of the de Cloiseaux-Pearson continuum\cite{des1962spin}, a characteristic of fractional spinon excitations. Upon increasing \(J_2\), the system enters a partially dimerized regime, where a spin gap opens and the DSF develops more structured intensity distributions. As \(J_2\) increases further, incommensurate features emerge, visible through the splitting of low-energy spectral peaks. Eventually, at large \(J_2\), the system enters a floating incommensurate phase, characterized by gapless excitations with incommensurate momenta, manifesting as a broad continuum in the DSF similar to the spin-$3/2$ chain.

To understand the nature of low-energy excitations in the partially dimerized phase, we carry out a SMA analysis on VBS states for both spinon and magnon modes, following the procedure established for the spin-\(3/2\) chain. As in that case, spinon excitations correspond to domain walls between two degenerate partially dimerized VBS states. We construct spinon dispersions and compare in Fig.~\ref{fig:10}a the two-spinon continuum (green shaded region, right panel) with the DSF data (left panel). The close match between the continuum and regions of significant spectral intensity strongly suggests that fractionalized spinon excitations dominate low-energy dynamics.

Magnon excitations, constructed by locally promoting a singlet bond into a triplet, are also analyzed within the SMA framework. The resulting magnon dispersion relations are shown as white lines in Fig.~\ref{fig:10}a. These modes traverse regions of high spectral weight, indicating their relevance as coherent, resonant excitations, even if embedded within a broader spinon continuum. As in the spin-\(3/2\) case, this coexistence underscores the complex interplay between fractional and collective excitations in higher-spin chains.

Figure~\ref{fig:10}b displays the calculated spinon dispersions, which exhibit clear incommensurate minima and nonmonotonic gap evolution with increasing \(J_2\). These features confirm the onset of incommensurate correlations within the partially dimerized phase and highlight the proximity to floating phase physics.

Together, the tDMRG and SMA results reveal a consistent picture: the spin-\(5/2\) \(J_1\)–\(J_2\) chain supports a partially dimerized intermediate phase, and its low-energy excitations are mainly fractionalized spinons. 

\section{Interpretation of the results}
\label{obsv}

\begin{comment}

\begin{figure*}[t]
    \centering
    \begin{tabular}{cc}
        % First Row
        \subfigure[]{
            \includegraphics[width=0.48\textwidth]{fig1.pdf}
            \label{fig:1}
        } &
        \subfigure[]{
            \includegraphics[width=0.48\textwidth]{fig3.pdf}
            \label{fig:3}
        } \\[-0.5em] % Reduce vertical space
        % Second Row
        \subfigure[]{
            \includegraphics[width=0.48\textwidth]{fig2.pdf}
            \label{fig:2}
        } &
        \subfigure[]{
            \includegraphics[width=0.48\textwidth]{fig4.pdf}
            \label{fig:4}
        } \\
    \end{tabular}
    \caption{(a) Gap of the two-spinon continuum as a function of \(J_2\). The gap reaches its maximum in the middle of the partially dimerized phase and decreases to zero near the phase edges.; (b, d) Comparison of wavevectors as functions of \(J_2\) for magnon and spinon continua in spin-\(3/2\) and spin-\(5/2\) chains, respectively. The plot highlights the transitions into incommensurate behavior for both excitations. (c) Two-spinon gap for the spin-\(5/2\) \(J_1\)-\(J_2\) Heisenberg chain. The maximum gap is observed near the middle of the partially dimerized phase, decreasing to zero at the boundaries.}
    \label{fig:grid}
\end{figure*}
\end{comment}

\begin{figure*}[t]
    \centering
    \begin{tabular}{cc}
        \subfigure[]{
            \includegraphics[width=0.48\textwidth]{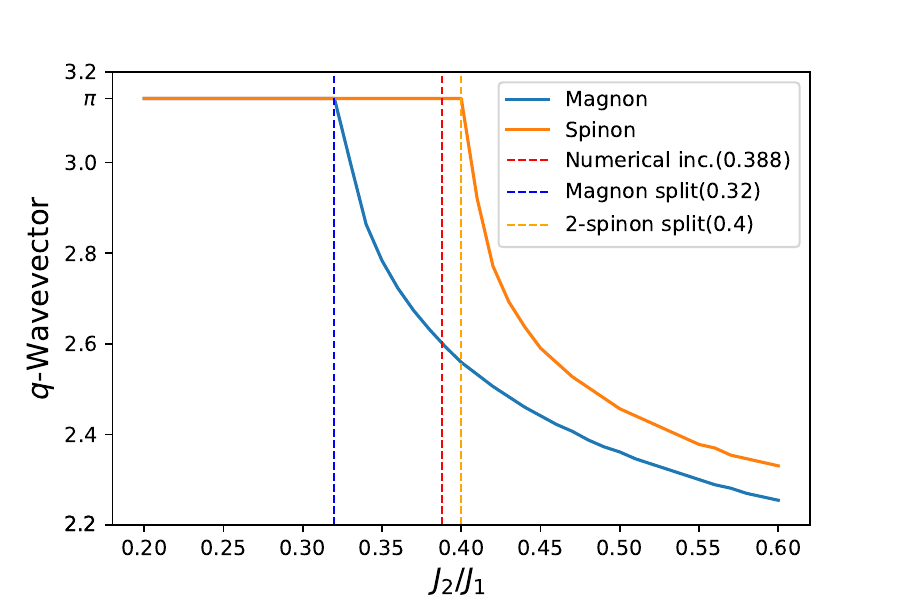}
            \label{}
        }&
        \subfigure[]{
            \includegraphics[width=0.48\textwidth]{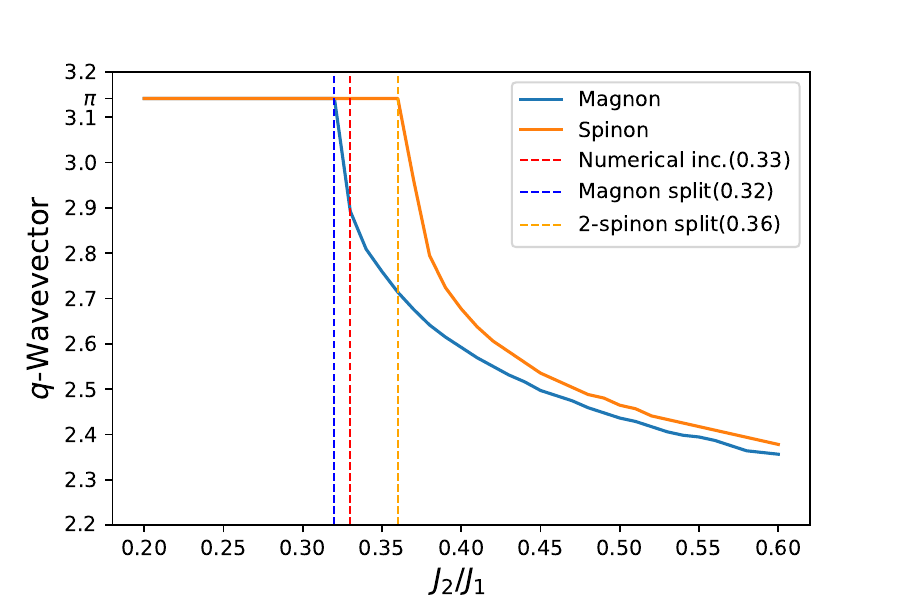}
            \label{}
        } \\
    \end{tabular}
    \caption{(a, b) Comparison of wavevectors as functions of \(J_2\) for magnon and spinon continua in spin-\(3/2\) and spin-\(5/2\) chains, respectively. The plot highlights the transitions into incommensurate behavior for both spin chains.}
    \label{fig:grid}
\end{figure*}

In this section, we summarize and interpret the key phenomenological trends observed in our numerical simulations and analytical calculations, with a particular focus on the evolution of spectral features across different coupling regimes in both spin-\(3/2\) and spin-\(5/2\) chains.

The transition into incommensurate correlations is captured through the wavevector evolution of both spinon and magnon dispersions. Fig.~\ref{fig:grid}a presents a comparison of wavevectors as functions of \(J_2\) for the magnon, and spinon continua, for spin-3/2 chain. The magnon dispersion becomes incommensurate at \(J_2 \approx 0.32J_1\). In contrast, the spinon dispersion becomes incommensurate at a later value, approximately \(J_2 \approx 0.4J_1\). According to Ref.~\cite{roth1998frustrated}, the Lifshitz point occurs at \(J_2 \approx 0.388J_1\), where the static structure factor develops two distinct peaks. This intermediate window, which spans the region around the Lifshitz point at \(J_2 \approx 0.388J_1\) highlights a subtle competition between two mechanisms of incommensuration: one driven by magnon propagation and the other by the underlying spinon domain wall excitations. Although magnons appear as well-defined modes in the SMA calculations, their signatures in the DSF suggest that they are not isolated quasiparticles. Rather, their dispersions traverse regions of high spectral weight within the broader spinon continuum, indicating that these magnons may be viewed as resonances—collective modes dressed by the surrounding fractional excitations. Indeed, the minimum of the magnon dispersion at an incommensurate value of the wave vector is reflected in the dispersion of a broad resonance, as can be seen in Fig.~\ref{fig:7}a(ii). This interpretation underscores the hybrid nature of excitations in frustrated systems, where the distinction between quasiparticles and continua becomes blurred by strong correlations.

A similar scenario emerges for the spin-5/2 chain, as illustrated in Fig.~\ref{fig:grid}b. Here, the magnon dispersion becomes incommensurate at $J_2 \approx 0.32J_1$, closely followed by numerical incommensuration at $J_2 \approx 0.33J_1$~\cite{chepiga20222}, and the spinon dispersion becomes incommensurate at a slightly higher coupling of $J_2 \approx 0.36J_1$. This indicates a universal pattern in higher-spin frustrated quantum chains, with quantitative differences emerging due to the reduced quantum fluctuations and altered interplay of spin excitations at higher spin magnitudes.

\begin{comment}
The evolution of the spinon gap as a function of the next-nearest neighbor interaction \(J_2\) reveals a non-monotonic behavior across the partially dimerized phase. The gap of the spinon dispersion initially increases and subsequently decreases, as illustrated in Fig.~\ref{fig:7}b. Fig.~\ref{fig:1} further shows the gap of the two-spinon continuum as a function of \(J_2\). The gap reaches its maximum in the middle of the numerically identified partially dimerized (PD) phase\cite{chepiga2020floating} and decreases to zero near the edges of this phase. Notably, the maximum gap for the two-spinon continuum occurs at \(J_2 = 0.42J_1\) , which is in close agreement with the numerically determined value of \(J_2 = 0.41J_1\)(in my own DMRG calculations). A similar trend is observed for the spin-\(5/2\) \(J_1\)-\(J_2\) Heisenberg chain, as shown in Fig.~\ref{fig:2}. For the spin-\(5/2\) case, the two-spinon gap again attains its maximum near the middle of the partially dimerized phase and reduces to zero at the phase boundaries. This dome-like structure is consistent with expectations from a valence bond solid (VBS) picture, where the singlet patterning is most robust in the central region of the PD phase.
\end{comment}

In both the spin-\(3/2\) and spin-\(5/2\) cases, when increading $J_2/J_1$ the spinon dispersion reaches zero with splitting, as illustrated in Fig.~\ref{fig:7}b,~\ref{fig:10}b. This behavior is indicative of a condensation of incommensurate spinons, heralding the emergence of the floating phase. 

Overall, the SMA framework, supported by high-precision tDMRG data, successfully disentangles the spectral landscape of the DSF into its elementary constituents, spinons, and magnons, thereby providing a detailed microscopic understanding of the rich dynamical behavior of the frustrated spin chains.

\section{Conclusion and discussion}
\label{concl}
In this work, we have performed a comprehensive study of the DSF of the frustrated spin-\(3/2\) and spin-\(5/2\) Heisenberg chains governed by the \(J_1\)-\(J_2\) model. Focusing on the partially dimerized phase and its adjacent transitions, we investigated the nature of low-lying excitations and their contributions to the DSF using a combination of numerical and analytical techniques. Leveraging the high accuracy of the DMRG and tDMRG methods, alongside analytical insight from the SMA, we identified the distinct roles of magnon and spinon excitations in shaping the observed spectral features.

Our results provide a deeper understanding of frustrated quantum magnetism, shedding light on the emergence of incommensurability in the DSF and its implications for experimental observations. Across both spin-\(3/2\) and spin-\(5/2\) cases, the spectral function is dominated by spinon continua, mirroring the spin-\(1/2\) case, and the magnon mode does not split out of the two-spinon continuum at any point in the Brillouin zone. This behavior contrasts with the spin-\(1/2\) Majumdar-Ghosh point \cite{majumdar1969next} at \(J_2/J_1 = 0.5\), where the magnon briefly detaches near \(k = \pi/2\). Furthermore, the SMA-derived spinon dispersion undergoes a transition to incommensurability as \(J_2\) increases. The gap of the spinon modes closes on both sides of the partially dimerized phase. At the lower boundary, the spinon mode condenses at the commensurate wave vector \(k = \pi/2\), leading to a standard Luttinger liquid phase with central charge $c=1$. In contrast, at the upper boundary, the condensation takes place simultaneously at two incommensurate wave vectors, leading to a DSF with two low-lying branches. This suggests that the central charge c=2, as it would be in the limit of decoupled chains $J_1=0$. Further numerical verification of this prediction, however, is extremely challenging due to incommensurate Friedel oscillations affecting the finite-size entanglement entropy profile. 

Furthermore our predictions for the DSF can be considered as predictions for the inelastic neutron scattering spectrum of frustrated spin chains. One compound is believed to be well described by the spin-\(5/2\) $J_1-J_2$ model with $J_2/J_1 \simeq 1$\cite{boya2021expt}. Our approach predicts that the spectrum should consist of a double Des Cloizeaux-Pearson continuum that touches zero energy at a slightly incommensurate wave-vector\cite{chepiga20222}. It would be interesting to check if other systems could be found that lie in the intermediate gapped region for which the DSF exhibits a richer structure.

On the methodological side, our use of SMA on both magnons and spinons has proven to be an effective tool in disentangling contributions to the DSF. While the magnons do not fully explain the DSF on their own, their dispersion curves coincide with regions of high spectral weight, suggesting that they may manifest themselves as resonances within the broader spinon continuum. This subtlety adds depth to the understanding of spectral mixing in frustrated spin systems, particularly in regimes with strong quantum fluctuations.

The framework we developed, combining VBS intuition with SMA and tDMRG, can be readily extended to other frustrated or topologically nontrivial spin systems. Future directions include the extension of this approach to the spin chains with three-site interactions, or two-dimensional analogs such as frustrated ladders and square lattices.

\begin{acknowledgments}
We thank Henrik R\o nnow for insightful discussions. A.S. acknowledges support from the Swiss Government Excellence Scholarship (FCS Grant No. 2021.0414). This work was supported by the Swiss National Science Foundation under Grant No. 212082. The calculations have been performed using the facilities of the Scientific IT and Application Support Center of EPFL.
\end{acknowledgments}

\bibliography{references.bib}

\appendix

\section*{APPENDIX}

\section{Estimation of Correlation Lengths}

In order to ensure that the chosen system sizes for the calculation of the dynamical structure factor are sufficiently large, we have estimated the spin--spin correlation lengths for a few parameters under investigation. The correlation lengths were extracted by fitting the equal-time spin--spin correlation function envelope to the Ornstein-Zernike form\cite{ornstein1918linearen,ornstein1926bemerkung}
\begin{equation}
    C_{i,j}^{\mathrm{OZ}} \propto \frac{e^{-|i-j|/\xi}}{\sqrt{|i-j|}},
\end{equation}
where $\xi$ denotes the correlation length. 

The numerical correlation data were fitted to this expression in logarithmic form, which linearizes the exponential decay and provides a more stable estimate for $\xi$. Only the envelope was considered in the fit, which effectively neglects the oscillatory contributions and isolates the exponential decay behavior.

\begin{figure}[t]
  \centering

  \subfigure{
    \includegraphics[width=\columnwidth]{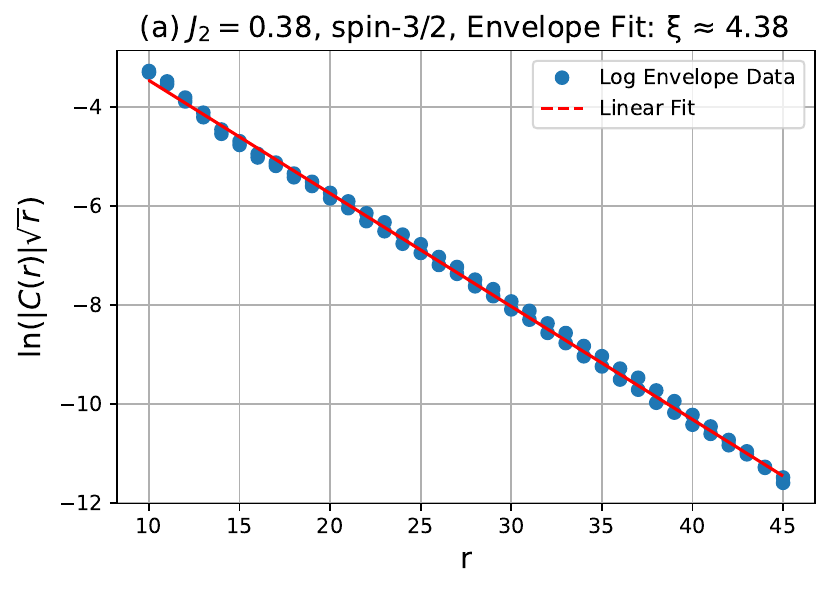}
    \label{fig:subfig_a}
  }\\[-1.2ex]

  \subfigure{
    \includegraphics[width=\columnwidth]{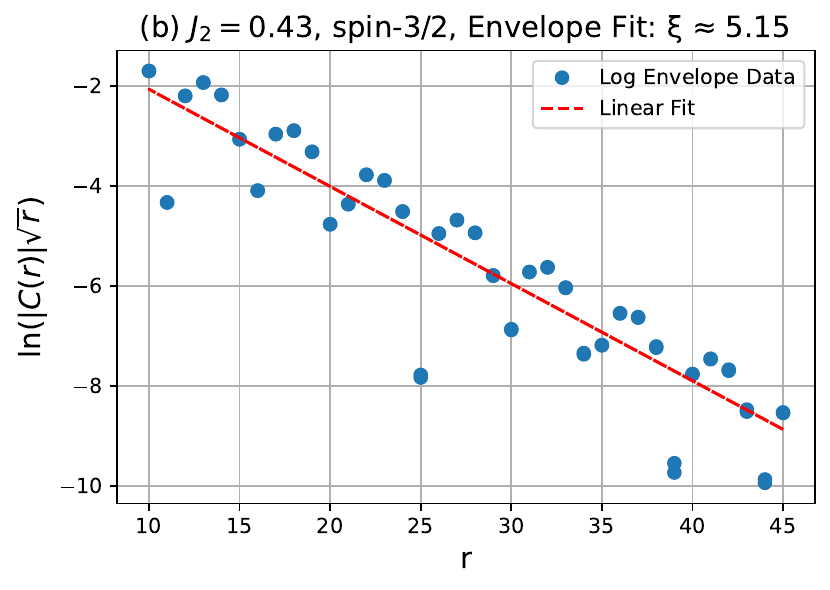}
    \label{fig:subfig_b}
  }\\[-1.2ex]

  \subfigure{
    \includegraphics[width=\columnwidth]{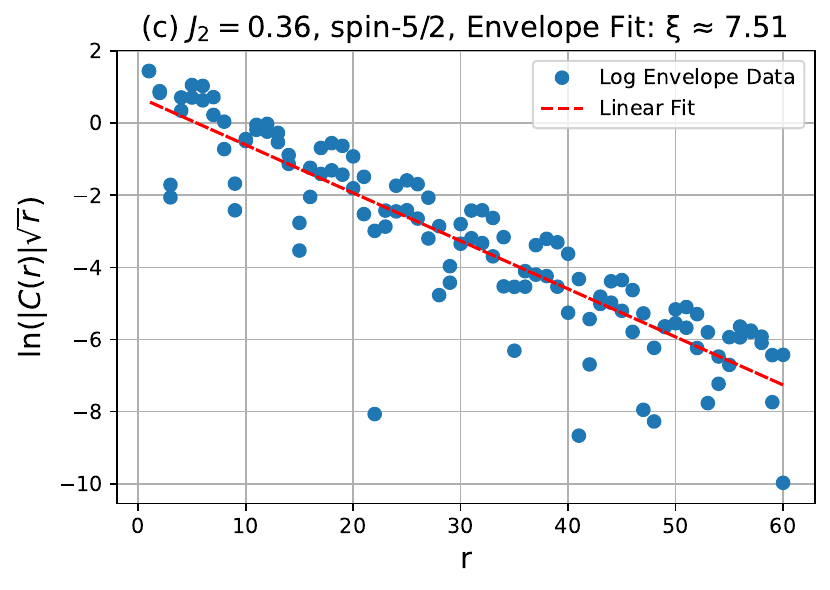}
    \label{fig:subfig_c}
  }

  \caption{Correlation envelope fits for different values of $J_2$ and spin magnitude.
           (a) $J_2 = 0.38$, spin-3/2; 
           (b) $J_2 = 0.43$, spin-3/2;
           (c) $J_2 = 0.36$, spin-5/2.}
  \label{envelop}
\end{figure}

For the spin-$3/2$ Heisenberg chain at $J_2 = 0.38$ and $J_2 = 0.43$, the extracted correlation lengths are $\xi \approx 4.38$ and $\xi \approx 5.15$, respectively(Fig. \ref{envelop}a,b). Similarly, for the spin-$5/2$ Heisenberg chain at $J_2 = 0.36$, we find a correlation length of $\xi \approx 7.51$(Fig. \ref{envelop}c). All these systems are located in gapped phases, and the finite values of $\xi$ confirm the presence of an excitation gap and short-range spin correlations. To estimate fitting uncertainties and validate the robustness of our procedure, we repeated the fits using only the local maxima (peaks) of the correlation function. When the number of oscillation periods is large, as it is in our system of \(L=150\) sites, both fitting procedures should converge to the same value. We indeed find close agreement: the peak-based fit yields \(\xi \approx 4.38\) and \(\xi \approx 5.11\) for spin-\(3/2\) at \(J_2 = 0.38\) and \(J_2 = 0.43\), respectively, and \(\xi \approx 7.22\) for spin-\(5/2\) at \(J_2 = 0.36\). The difference between these two estimates can be interpreted as a rough estimate of the fitting uncertainty for a given system size.

The system sizes employed for the dynamical structure factor calculations were chosen to be significantly larger than these correlation lengths, ensuring that finite-size effects are well-controlled and the extracted spectra are physically meaningful.

\section{Convergence with Bond Dimension in DMRG Calculations}
\label{convergence}

\begin{figure}[t]
  \centering

  \subfigure[]{
    \includegraphics[width=\columnwidth]{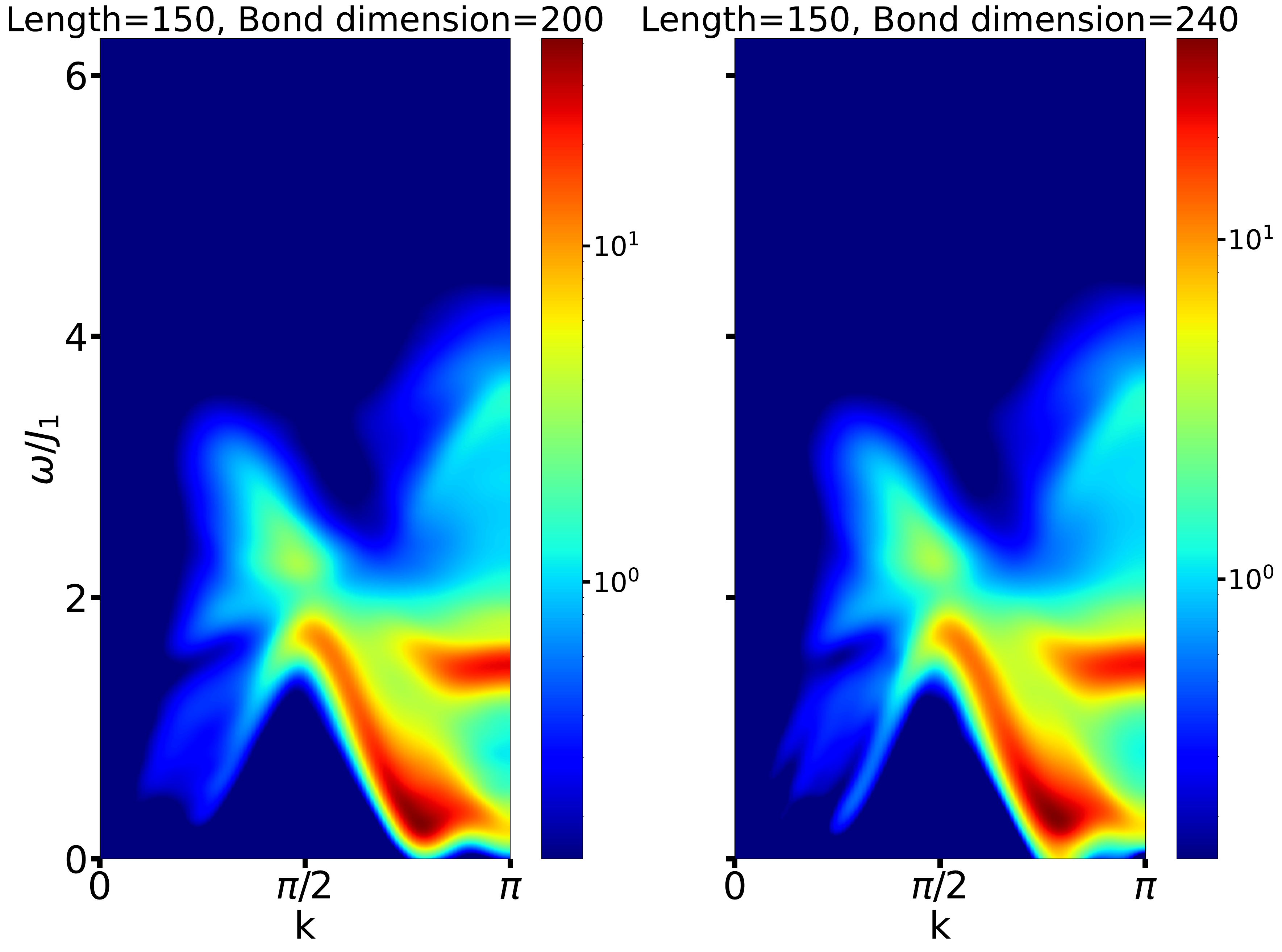}
    \label{fig:subfig_a}
  }\\[-1.2ex]

  \subfigure[]{
    \includegraphics[width=\columnwidth]{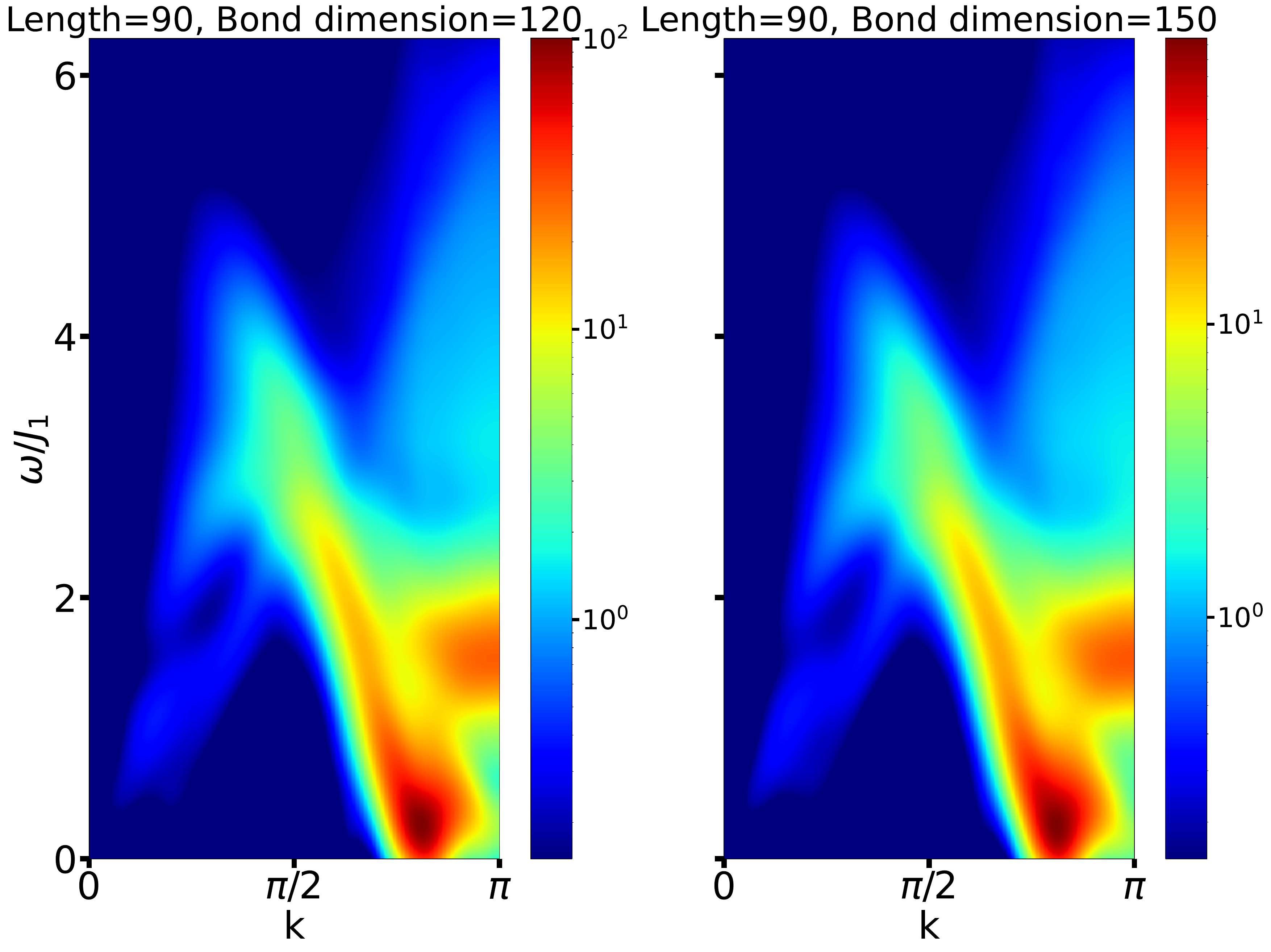}
    \label{fig:subfig_c}
  }

    \subfigure[]{
    \includegraphics[width=\columnwidth]{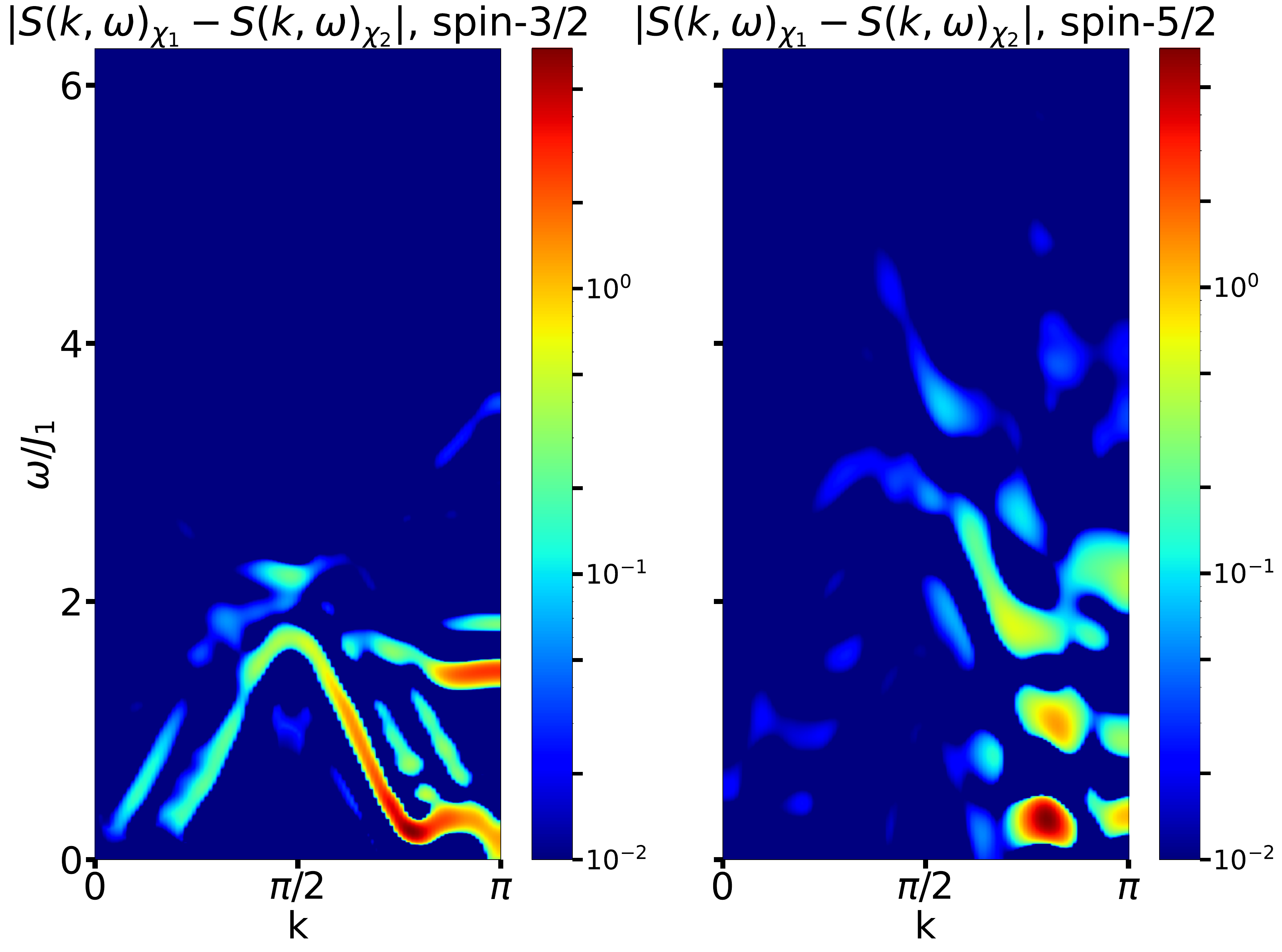}
    \label{fig:subfig_c}
  }

  \caption{(a) Spin-$3/2$ \(J_1\)–\(J_2\) chain at \(J_2 = 0.43J_1\), computed for a chain of length \(L = 150\) with bond dimensions \(\chi = 200\) and \(\chi = 240\). 
(b) Spin-$5/2$ \(J_1\)–\(J_2\) chain at \(J_2 = 0.35J_1\), for a chain of length \(L = 90\), with \(\chi = 120\) and \(\chi = 150\). In both cases, the DSF spectra remain qualitatively and quantitatively consistent, confirming that the chosen bond dimensions are sufficient to capture the essential features of the excitations. (c) Absolute difference \(|S(k,\omega)_{\chi_1} - S(k,\omega)_{\chi_2}|\) between the two bond dimension results shown in (a) and (b), respectively.}
  \label{BDs}
\end{figure}

To verify the reliability of the DSF results presented in this work, we have examined the potential influence of the fixed bond dimension used during the ground state optimization. While the dynamical correlation functions are computed using a MPS framework, the quality of the ground state approximation is essential, as it directly influences the computed spectral properties.

To rule out the possibility that the system is trapped in a metastable or locally suboptimal MPS due to an insufficient bond dimension, we have performed comparative calculations at two different system sizes and two corresponding bond dimensions for selected representative cases.

Specifically, for the spin-$3/2$ Heisenberg chain at $J_2 = 0.43$, we have carried out two independent sets of simulations: one at a system size of $L = 150$ with a fixed bond dimension $\chi = 200$, and another at $L = 150$ with a bond dimension $\chi = 240$(Fig. \ref{BDs}a). In both cases, the computed dynamical structure factors exhibit the same spectral features, with no qualitative or significant quantitative deviations observed. Similarly, for the spin-$5/2$ Heisenberg chain at $J_2 = 0.35$, we have performed calculations at $L = 90$ with $\chi = 120$, and at $L = 90$ with $\chi = 150$(Fig. \ref{BDs}b). Again, the spectra remain consistent across the two bond dimensions, confirming that the observed features are not artifacts of a limited variational space, but are genuine characteristics of the system.

To better visualize the effect of bond dimension on the computed DSF spectra, we also present in Fig.~\ref{BDs}(c) the absolute difference between spectra obtained at two different bond dimensions: \(|S(k,\omega)_{\chi_1} - S(k,\omega)_{\chi_2}|\), for both spin-\(3/2\) and spin-\(5/2\) chains. This difference plot directly quantifies the impact of varying \(\chi\), and reveals that the changes are negligible.

These tests demonstrate that the fixed bond dimension employed throughout the ground state calculations is sufficient for the parameter regimes studied here and does not compromise the physical reliability of the dynamical structure factor results presented in this work.

\end{document}